\documentclass[10pt,final,journal,letterpaper,twoside,twocolumn]{IEEEtran}

\usepackage{cite}
\usepackage{graphicx,color,epsfig,rotating}
\usepackage{amsfonts,amsmath,amssymb,bbm}
\usepackage{algpseudocode}
\usepackage{algorithm}
\usepackage{amsmath}
\usepackage{cite}
\usepackage{mdwtab}
\usepackage{placeins}
\usepackage{psfrag, graphicx}
\usepackage[latin1]{inputenc}
\usepackage{amssymb}
\usepackage{makeidx}
\usepackage{xcolor}
\usepackage{booktabs}
\usepackage{subcaption}
\usepackage{url}

\long\def\comment#1{}

\newfont{\bbb}{msbm10 scaled 700}

\newfont{\bb}{msbm10 scaled 1100}

\newcommand{\Xc}{{\cal X}}

\newcommand{\eqdef}{\stackrel{\Delta}{=}}

\newcommand{\be}{\begin{equation}}
\newcommand{\ee}{\end{equation}}
\newcommand{\bea}{\begin{eqnarray}}
\newcommand{\eea}{\end{eqnarray}}

\newtheorem{defn}{Definition}
\newtheorem{example}{Example}
\newtheorem{theorem}{Theorem}
\newtheorem{corollary}{Corollary}
\newtheorem{proposition}{Proposition}
\newtheorem{remark}{Remark}

\begin{document}

\setcounter{page}{1}

\title{FLCD: A Flexible Low Complexity Design of Coded Distributed Computing}

\author{Nicholas Woolsey,~\IEEEmembership{Student Member,~IEEE}, Xingyue Wang, \\ Rong-Rong Chen,~\IEEEmembership{Member,~IEEE}, and Mingyue Ji,~\IEEEmembership{Member,~IEEE}
\thanks{The authors are with the Department of Electrical Engineering,
University of Utah, Salt Lake City, UT 84112, USA. (e-mail: nicholas.woolsey@utah.edu, u1093573@utah.edu,  rchen@ece.utah.edu and mingyue.ji@utah.edu) }
}

\maketitle



\begin{abstract}
We propose a flexible low complexity design (FLCD)   of coded distributed computing (CDC) with empirical evaluation on Amazon Elastic Compute Cloud (Amazon EC2).
CDC can expedite MapReduce like computation by trading increased map computations to reduce communication load and shuffle time.
\if{0}
Since a CDC scheme requires strategic file mapping, shuffle, and function assignment  which may have high complexity or  limited flexibility, 
it is challenging to design effective schemes  amenable for practical implementation in most parameter regimes.
\fi
A main novelty of FLCD 
is to utilize the design freedom in defining map and reduce functions to develop asymptotic homogeneous systems  to support varying
intermediate values (IV)  sizes under a general MapReduce framework. 
Compared to existing designs with constant IV sizes, 
FLCD offers greater flexibility in adapting to network parameters 
and 
significantly reduces the implementation complexity by requiring fewer input files and shuffle groups. 
The FLCD scheme  
is the first proposed low-complexity CDC design that can operate on a network with an arbitrary number of nodes and computation load.
We perform empirical evaluations of 
the FLCD 
by executing 
the TeraSort algorithm on an Amazon EC2 cluster. {This is the first time that theoretical 
predictions of the CDC shuffle time 
are validated by empirical evaluations.}
The evaluations demonstrate a 2.0 to 4.24$\times$ speedup compared to conventional uncoded MapReduce,  
a 12$\%$ to 52$\%$ reduction in total time, 
and a wider range of operating network parameters compared to existing CDC schemes. 
\end{abstract}


\section{Introduction}
\label{sec: intro}

Communication load is considered as one of the major bottlenecks in current distributed computing platforms such as MapReduce \cite{dean2008mapreduce} and Spark \cite{zaharia2010spark}.
Coded distributed computing (CDC) introduced by Li, Maddah-ali, Yu and Avestimehr (LMYA) in \cite{li2017fundamental} is an efficient approach to execute complex queries on massive databases over such systems while
minimizing the use of communication resources. 
In this paper, we consider a network with $K$ computing/worker nodes, where each node is assigned to compute a set of output functions, each of which is a function of all $N$ input files. Each file is available at some subset of nodes. The computing process is divided into Map and Reduce phases. In the Map phase, each node computes an intermediate value (IV) for each output function from each locally available file. 
In the Shuffle phase, each node transmits the needed IVs for other computing nodes such that all the nodes can reduce the assigned output functions by jointly processing the $N$ IVs. We define the computation load $r$ as the total number of times that each IV is computed in the system. By using the novel LMYA design \cite{li2017fundamental},  the optimal communication load $L$, defined as the total number of bits transmitted among nodes in the shuffle phase normalized by the size of all IVs, is given by $L^{\rm LMYA}=\frac{1}{r}\left(1-\frac{r}{K}\right)$. This is a surprising result since the gain of the computation load $r$ is multiplicative, i.e., if $r$ is doubled, then $L$ is roughly halved. Due to this significant theoretical advantage of CDC, it has attracted enormous attention recently \cite{li2016unifiedstraggler,8647133,li2018compressed,kiamari2017Globecom,Ezzeldin2017alternative,7935426,woolsey2018new,woolsey2019cascaded,woolsey2019coded,Srinivasavaradhan2018random,8849227,xu2019heterogeneous,wan2020topological,Konstantinidis2020,jiang2020coded}.

Despite the attractive theoretical result of the LMYA design, several drawbacks prevent it from practical implementations. First, in order to achieve the promised multiplicative gain, it needs ${K \choose r} \approx 2^{K H_b(r/K)}$ input files, where $H_b(\cdot)$ is the binary entropy function. This means that the number of input files grows exponentially with the number of users $K$ for a fixed $H_b(r/K)$. Hence, the LMYA scheme is not  feasible for large $K$. Second, the key idea to achieve the multiplicative gain is to use {\em coded multicast} introduced in the index coding and coded caching literatures \cite{bar2011index,maddah2014fundamental,maddah2015decentralized,wan2020fundamental,ji2016fundamental}, where each coded multicast message is simultaneously useful for a shuffle group of multiple nodes instead of just a single node as in conventional unicast. Since LMYA requires ${K \choose r+1} \approx  2^{K H_b((r+1)/K)}$  shuffle groups to achieve the promised gain, the  large overhead for building these shuffle groups induces prohibitive complexity for practical implementation as first observed in \cite{li2017fundamental}. Third, the LMYA design strictly requires that 
all IVs have the same size such that the computation loads for different workers are exactly the same.
This strictly limits the flexibility of the CDC design since allowing sightly different computation loads among workers may not be very harmful in practice. In addition, requiring the same computation loads may be violated in practice due to the heterogeneous nature of distributed computing systems.
We will show in this paper that if this restriction is relaxed slightly, meaning that the computation loads among workers are approximately the same as $K/r$ becomes large, surprisingly, the complexity of the proposed CDC scheme can be reduced significantly. Meanwhile, the achieved communication load can be even smaller than  $L^{\rm LMYA}$.
Due to these drawbacks, the LMYA design has only been implemented when $r\leq 5$ under some parameter ranges in order to see gains compared to the conventional unicast shuffle scheme. In \cite{8647133,Konstantinidis2020}, Konstantinidis and Ramamoorthy (KR) proposed another implementable scheme based on a resolvable design to reduce the complexity of CDC and showed the gain of this scheme compared to the LMYA design via practical evaluation on Amazon EC2. However, the KR design is not flexible because it requires $K/r$ to be a positive integer. 
In \cite{kiamari2017Globecom,xu2019heterogeneous}, the heterogeneous system was investigated. However, the optimal designs have not been empirically tested and in theory only work for a system with $3$ nodes or have a complexity comparable to the LMYA design.

In this paper, we propose a flexible, low complexity design (FLCD)
to overcome the aforementioned limitations of existing approaches. The key novelty of FLCD is to slightly relax the strict requirement that each worker has the same local computation load  by allowing varying IV sizes such that any integer computation load $r\in\{ 2,\ldots, \frac{K}{2}\}$ is achievable and FLCD can be implemented for large $r$.
To the best of our knowledge, the proposed FLCD
is the first implementable scheme in the literature to address the limitations of CDC mentioned above.
Our contribution are summarized as follows.
\begin{itemize}
\item 
We propose 
 FLCD for CDC that significantly reduces the required numbers of input files and shuffle groups compared to the LMYA scheme. {
This leads to a more flexible design that not only works for a wider range of system parameters but also facilitates low complexity implementation.} { The FLCD scheme
is the first proposed low-complexity CDC design that can operate on a network with an arbitrary number of nodes and computation load. }
\item 
 We are the first to investigate the impact of varying IV size in CDC,  motivated by the fact that for many applications, relative IV size can be a useful design parameter.
The FLCD scheme allows slightly different IV sizes while keeping the computation loads of different workers approximately the same.\footnote{It means that the computation loads among workers converge to the same value as $K/r \rightarrow \infty$.} We call such systems as asymptotic homogeneous systems. 
It leads to the surprising result that 
the achievable communication load by FLCD can be lower than that of LMYA scheme, i.e.,  the previously established fundamental limit is ``breakable" by 
implementing  asymptotic homogeneous systems
{ with carefully designed IV sizes. 
To the best of our knowledge, this is the first work to examine the impact of varying IV sizes on 
the fundamental limit of CDC network. }
This opens up a new research direction for designing novel CDC schemes for such asymptotic homogeneous systems
for which the fundamental limits  remain unknown.
\item 
Instead of restricting $K/r$ to be an integer, the FLCD scheme is  applicable for any $K/r$, when $2 \leq r \leq K/2$, while maintaining  the multiplicative gain of the CDC at a greatly reduced complexity. 
The key idea to achieve this is to allow flexible IV sizes which can be approximately the same when $K/r$ becomes large.
\item We demonstrate the effectiveness of the proposed FLCD scheme  using sorting, which is a fundamental building block of many machine learning algorithms. In particular, we use the benchmark of TeraSort and implement it via FLCD on Amazon EC2. The evaluations demonstrate a 12$\%$ to 52$\%$ reduction in total  time (this includes not only the shuffle time, but also five other important time metric)  compared to LMYA and KR schemes, a 2.0 to 4.24$\times$ speed-up compared to conventional uncoded MapReduce,
{ and confirms the greater flexibility of FLCD  under varying node numbers and computation requirement of  CDC networks.}
{ This is also the first time that theoretical predictions of the shuffle time of a CDC design are validated by empirical evaluations.}
\end{itemize}

While  the proposed FLCD schemes in this work originate from previously developed combinatorial designs for CDC networks \cite{woolsey2020new,woolsey2020cascaded}, a key difference is that FLCD leverages the design freedom in defining map and reduce functions to support  varying IV sizes in a more general MapReduce framework. Compared to
\cite{woolsey2020new,woolsey2020cascaded} which focus on heterogeneous systems, 
this new approach puts  a different emphasis on asymptotic homogeneous systems and aims to design more flexible CDC schemes   that can operate under a wider range of system parameters. A unique contribution of this work is the successful validation of the FLCD through empirical evaluations on AMAZON EC2. This provides strong evidence on the effectiveness of the combinatorial designs utilized in not only this work, but also those in \cite{woolsey2020new,woolsey2020cascaded}. 

This paper is organized as follows. In Section~\ref{sec: model}, we present the proposed system model with varying IV sizes. This is followed by a description of  the state-of-the-art achievable designs with fixed IV sizes in Section \ref{sec:state-of-art}. We then present an example in Section \ref{sec: examples} to illustrate the basic ideas of the proposed FLCD scheme.  The main results of the FLCD scheme in terms of communication-computation tradeoff and complexity are given in Section~\ref{sec: Main Results}.
In Section~\ref{sec: FLCD}, we provide the general design of the FLCD scheme and present a detailed example.    In Section~\ref{sec: empirical evaluation}, we present and discuss empirical system evaluations of FLCD on Amazon EC2. Concluding remarks are given in Section~\ref{sec: conclusions}. Key proofs are given in the Appendices.


\paragraph*{Notation Convention}
We use $|\cdot|$ to represent the cardinality of a set or the length of a vector. 
Also $[n] := [1,2,\ldots,n]$ for some $n\in\mathbb{Z}^+$, where $\mathbb{Z}^+$ is the set of all positive integers, and $\oplus$ represents bit-wise XOR. 
Furthermore, let $\mathbb{R}$ be the set of all real numbers. The notation $\lfloor m \rfloor$ means the floor operator of $m$.


\section{Proposed System Model with Varying IV size} 
\label{sec: model}
The considered system model is motivated by the fact that the relative size of IVs is a design choice for many MapReduce applications.
The goal of this work is to utilize the design freedom in defining map and reduce functions to develop new CDC designs to support varying IV sizes in order to reap additional benefits in terms of communication load, overall computation time and complexity. 
This is in contrast to previous CDC designs (e.g., \cite{li2017fundamental,7935426,Konstantinidis2020}) that can only operate with constant IV sizes.
Specifically, we adopt a more general system model of the original CDC work 
\cite{li2017fundamental}. We consider a network of $K$ nodes, labeled $1,\ldots,K$. The whole dataset is split into $N$ equally sized input files, $\{w_1,\ldots,w_N\}$, based on the specific 
design.
The system aims to compute $K$ output functions, $\phi_k(w_1,\ldots,w_N), k \in [K]$, each of which requires all $N$ files as input.
The output function $\phi_k$ is assigned to node $k$.\footnote{Each output function can be a collection of many functions. When there are more functions than nodes, we can group the functions into $K$ non-overlapping sets and define these sets as output functions.}
In general, as 
the dataset may be large, 
node $k$ only has access to a subset of files  $\mathcal{M}_k\subseteq\{ w_1,\ldots,w_N\}$ and each file is available at $r$ nodes.
Since nodes do not have access to all $N$ files such that they cannot directly compute their assigned output function, a MapReduce framework is used which has three phases as follows. { While in a typical MapReduce framework, the IV sizes are fixed,  in the following, we describe a MapReduce framework that allows varying IV sizes. The proposed FLCD scheme is based on this modified framework. }


{\bf \em Map Phase}: Nodes compute IVs from locally available files. Each node $k$ uses each locally available file $w_n\in\mathcal{M}_k$ as input to the {\it map functions}, $\{g_{1,n},\ldots, g_{K,n}\}$, to compute the IVs, $\{v_{1,n},\ldots,v_{K,n}\}$, with possibly different lengths, i.e., $v_{k,n} = g_{k,n}(w_n)$ and $|v_{k,n}|=T_k$ bits. The relative IV sizes, $T_k, k \in [K]$, are based on the choice of the specific map and reduce function designs.

{\bf \em Shuffle Phase}: Each node $k$ broadcasts a (coded) message set $\mathcal{X}_k$ on a shared-link, over which $\mathcal{X}_k$ sent by node $k$ can be received by all other nodes without errors. 
All (coded) messages, $\mathcal{X}_k$, are designed so that each node $k$ can collect every IV $v_{k,n}$ for $n\in [N]$. In general, messages, $\mathcal{X}_k$, may include coded combinations of IVs such that nodes can decode requested IVs using locally computed IVs.

{\bf \em Reduce Phase}: Each node $k$ computes the {\it reduce function}, $h_k(v_{k,1},\ldots,v_{k,N})$, with all IVs, $\{v_{k,1},\ldots,v_{k,N}\}$, as input. The map and reduce functions are designed such that $h_k(v_{k,1},\ldots,v_{k,N}) = \phi_k(w_1, \cdots, w_N)$.

Under this framework we define the {\it computation load}, $r$, as the mean number of times each file is mapped to the 
system where
\begin{align}
r\triangleq\frac{1}{N}\sum_{k=1}^{K}|\mathcal{M}_k|,
\end{align}
which can be understood as the number of times that each IV is computed in the system.
In conventional uncoded MapReduce we find $r=1$ where each file is only mapped once on the computing network.
In this paper, for simplicity, we will just consider the case when $r$ is an integer.\footnote{The case of non-integer $r$ can be solved by using the similar memory-sharing scheme used in classical coded caching literature \cite{maddah2014fundamental}.}
Next, we will present an example of TeraSort to illustrate the the system model described above and put particular focus on the heterogeneous IV sizes. TeraSort is widely used as a benchmark for MapReduce platform. 
\begin{example}\label{example: TeraSortFunctions} {\it TeraSort Map and Reduce Function Design}: The $K$ computing nodes aim to use TeraSort to sort a large set of integers in the range of $[0,Z)$  in a distributed manner. We design $K$ reduce functions, where node $k$ is assigned reduce function $h_k, \forall k \in [K]$. The reduce functions sort integers of a specific range defined by bounds $z_0,z_1,\ldots,z_K$ in an ascending order with $z_0=0$ and $z_K=Z > 0$. Reduce function $h_k$, sorts integers  in the range of $[z_{k-1},z_k)$. Then, map functions are designed to hash the integers of each file into bins defined by the bounds. Map function $g_{k,n}$ returns the IV $v_{k,n}$ which includes the integers of file $w_n$ in the range of $[z_{k-1},z_k)$. After the map phase, the nodes shuffle the corresponding IVs such that each node $k$ collects all integers in range $[z_{k-1},z_k)$ to be sorted with reduce function $h_k$. After the reduce phase, the integers will be sorted across the computing network. \hfill $\triangle$
\end{example}

Compared to previous works in CDC, we study a more general framework where IVs can have varying sizes that are dictated by the map and reduce function designs. Let the number of bits of IV $v_{k,n}$ be $T_k$ bits, which only depends on the corresponding reduce function, $h_k$. 
Then, we define the {\it communication load}, $L$, as the number of bits transmitted on the shared-link normalized by the total number of bits from all IVs
\begin{align}
  L\triangleq\frac{\sum_{k=1}^{K}|\mathcal{X}_k|}{N\sum_{k=1}^{K}T_k},
\end{align}
where $|\mathcal{X}_k|$ is the total number of bits from all transmitted messages in $\mathcal{X}_k$.
\begin{defn}
The optimal communication load or the optimal communication-computation tradeoff is defined as 
\be
L^*(r) \eqdef \inf\{L: (r, L) \text{ is feasible}\}.
\ee
\hfill $\Diamond$
\end{defn}
Next, we will present an example based on the previous TeraSort example to illustrate 
the possibility to vary the sizes of IVs in practice.

\begin{example}\label{example: varyingIVs}
{\it Design Choice of Relative IV Sizes}: The design of the map and reduce functions dictates the sizes of the IVs. Continuing Example \ref{example: TeraSortFunctions}, assuming the $N$ files are the same size and the integers follow a uniform distribution, the 
size of IV $v_{k,n}$ is $T_k = \frac{z_k-z_{k-1}}{Z}\cdot|w_n|$ with high probability where $|w_n|$ is the size in bits of each file. The bounds, $z_0,z_1,\ldots,z_K$, can be chosen accordingly to have desired varying IV sizes.\hfill $\triangle$
\end{example}


In contrast to the proposed system model that supports varying IV sizes, the state-of-the-art CDC designs typically assume constant IV sizes.
We will provide a brief description of these as below.

\section{State-of-the-Art Achievable Designs with Fixed IV sizes}
\label{sec:state-of-art}
Currently, there are only two CDC designs whose performance has been demonstrated through empirical evaluations over Amazon EC2 as shown in \cite{li2017coded,li2017fundamental,8647133,Konstantinidis2020}. These both assume constant IV sizes. The first is the LMYA design of \cite{li2017fundamental,li2017coded}. Under the system model in \cite{li2017fundamental}, the LMYA design achieves the information theoretic optimal communication-computation load tradeoff of
\begin{align}
L^{\rm LMYA}(r) = L^*(r) =\frac{1}{r}\left(1 - \frac{r}{K} \right).
\label{eq:fund_limit}
\end{align}
We will show in this paper that, this tradeoff is only optimal under the specific design framework of \cite{li2017fundamental}, which assumes the same IV sizes across the network.
In the LMYA design, each message of $\Xc_k$ sent from node $k$ is a network coded multicast message that  
serves $r$ independent requests from a group of any $r$ nodes simultaneously. Each of the $r$ nodes in such an multicast group can successfully decode requested IVs from the coded multicast message of $\Xc_k$. Hence, it can be seen that a multiplicative gain of $r$ in terms of communication load can be achieved using this coded multicasting scheme compared to the conventional unicast approach.
Although the promising theoretical performance achieved by the LMYA design, it has a high complexity because it requires $N = {K \choose r}$ input files, where each file is mapped to a set of $r$ nodes.
Also, the scheme requires $G={K\choose r+1}$ shuffle/multicast groups, significantly increasing the overhead in actual CDC implementations as shown in \cite{li2017fundamental,li2017coded}.

The second practically implemented design is the KR design of \cite{8647133,Konstantinidis2020} which has a reduced complexity as it only requires $N=\left(\frac{K}{r}\right)^{r-1}$ files and $G=\left(\frac{K}{r}\right)^{r-1}\left( \frac{K}{r} - 1 \right)$ shuffle groups. Both files and shuffle groups have been reduced exponentially 
compared to those of the LMYA design. Moreover, the multiplicative gain of coded multicasting is maintained and the communication load is
\begin{align}\label{eq: L_KR}
L^{\rm KR}(r) = \frac{1}{r-1}\left(1 - \frac{r}{K} \right),
\end{align}
which is asymptotically optimal as $r$ goes to infinity.
However, this scheme only works for homogeneous systems (i.e., the size of all $\mathcal{M}_k$ are the same)
and only holds for the limited parameter settings where
$m=\frac{K}{r}$ is an integer. This requirement can be very restrictive. For example, if $K=10$, then the possible choices of $r$ are only  $1, 2, 5, 10$, where $r=1$ (conventional MapReduce system) and $r=10$ (no needed communication) are not interesting cases.
Moreover, both the LMYA and KR designs only operate under the assumption of homogeneous IV sizes.

\section{Examples of the Proposed FLCD for $K=3$} 
\label{sec: examples}

In this section, we present two examples to illustrate the key idea of the proposed FLCD scheme 
for the special case $K=3$ nodes.  Although the specific designs described here  are different from those of the general design for $K>3$,  these examples outline the fundamental concepts of our system model and demonstrate how to design the Shuffle phase when IVs have varying size.
Moreover, the  examples demonstrate that under our general design framework which allows different IV sizes, the fundamental tradeoff, $L^{\rm LMYA}$ of \cite{li2017fundamental} originally derived for the homogeneous IV sizes no longer holds.
Example 3 outlines the conventional uncoded MapReduce approach based on unicast where each input file is mapped at exactly $r=1$ node.  Even for this uncoded case, we show that allowing variable IV sizes results in a lower communication load than that of \cite{li2017fundamental}. In Example 4, FLCD is applied to a network of $K=3$ nodes and each input file is mapped to $r=2$ nodes. This example uses coded multicasting, and our design with varying IV sizes again improves on the communication load of \cite{li2017fundamental}.



 \begin{figure}
  \centering
  \includegraphics[width=2.7in]{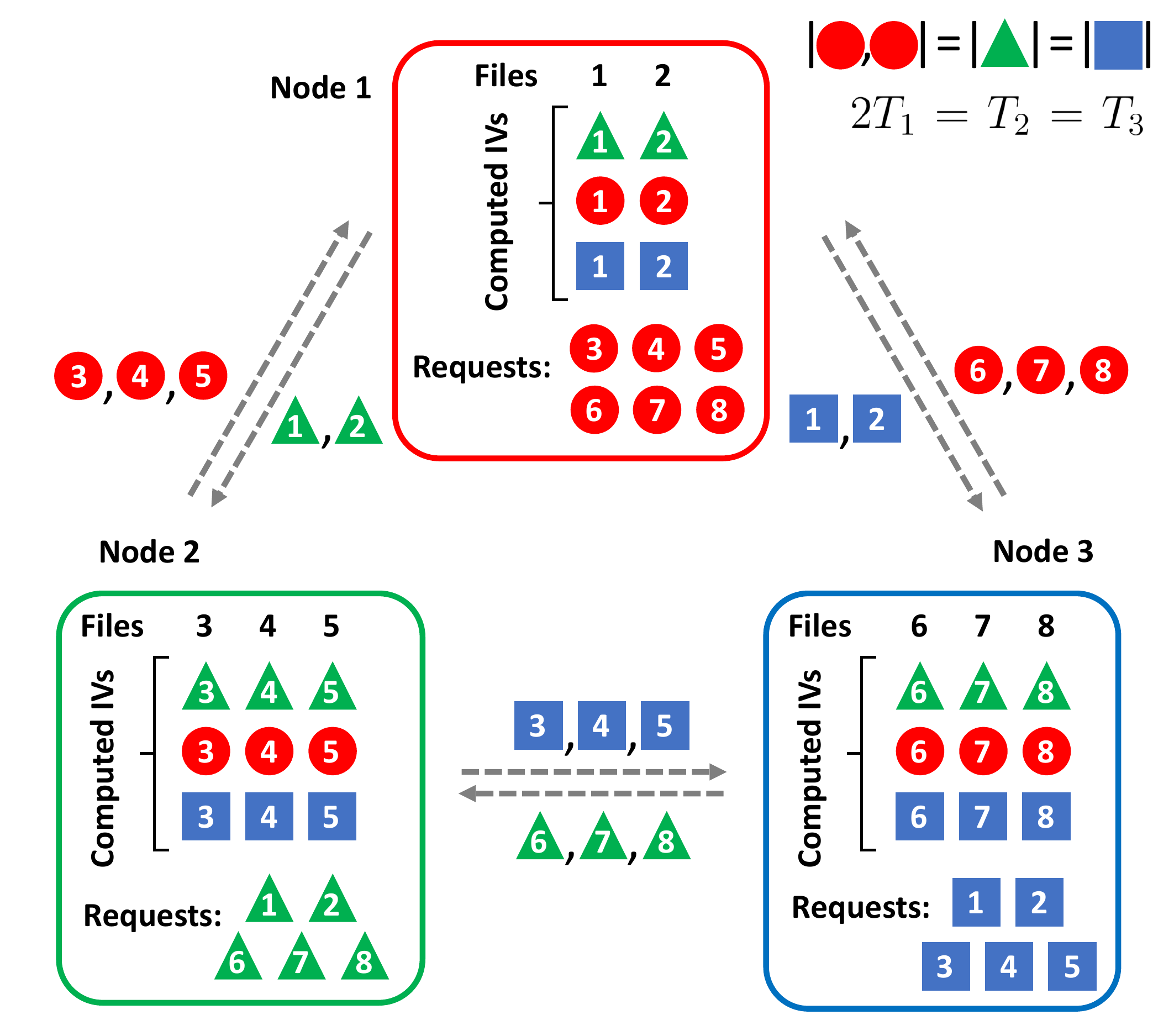} 
  \caption{  
  An example of conventional uncoded MapReduce with $r=1$, $N=8$ files and $K=3$ nodes. Node $i$ is assigned reduce function $h_i$. Three sets of 8 IVs,
    $\{v_{i,j}, j \in [8]\}$, corresponding to $h_i$, are represented by  red circles,  green triangles, and  blue blocks, respectively, for $i=1,2,3$,  with  file index $j$ labeled in the center. 
  Since files $1$ and $2$ are mapped to node $1$, it computes 6 IVs from these two files, indicated by ``computed IVs''. 
  These include two green triangles $v_{21}, v_{2,2}$, two red circles $v_{1,1}, v_{1,2}$, and two blue blocks  $v_{3,1}, v_{3,2}$. Since node $1$ is assigned $h_1$, it will request 6 IVs, each from a file that it does not have, shown as the 6 red circles  $\{v_{1,j}, j=3,4, \cdots, 8\}$.  Node $1$ sends two green triangles $v_{2,1}, v_{2,2}$ to node $2$ and two blue blocks  $v_{3,1}, v_{3,2}$ to node $3$ to fulfill their requests. Node $1$ also receives $3$ red circles $v_{1,3}, v_{1,4}, v_{1,5}$ from node $2$ and another set of $3$ red circles $v_{1,6}, v_{1,7}, v_{1,8}$ from node $3$ to meet its own requests. The IVs associated with 
  nodes $2$ and $3$ are defined similarly.}
  \label{fig: uncoded_example}
\end{figure}


\begin{example}\label{example: uncoded}{\it Conventional Uncoded MapReduce}: As shown in  Fig.~\ref{fig: uncoded_example}, a network of $K=3$ nodes aims to compute $3$ output functions,  one  assigned to each node. There are $N=8$ input files and each is mapped to  $r=1$ computing node.   Nodes $1$, $2$ and $3$ map the files of $\mathcal{M}_1=\{ w_1, w_2 \}$, $\mathcal{M}_2=\{ w_3, w_4, w_5 \}$ and $\mathcal{M}_3=\{ w_6, w_7, w_8 \}$, respectively. In the map phase, each node computes $3$ IVs, one for each output function, from each of its locally available files. Moreover, the map functions are designed such that $2T_1=T_2=T_3$ and  IVs for node $1$'s reduce function (red circles) contain half the number of bits of IVs for node $2$
 and $3$'s reduce functions (green triangles and blue squares, respectively). 

The shuffle phase is necessary so each computing node can collect the needed (or requested) IVs for its reduce function corresponding to its assigned output function. Nodes $1$, $2$ and $3$ will be required to have the access to all the $8$ IVs represented by red circles, green triangles, and blue squares, respectively, as shown in Fig. \ref{fig: uncoded_example}. In order to accomplish this, each node transmits the required IVs to the other two nodes on the shared-link. For instance, node $1$ transmits  IVs $v_{2,1}$ and $v_{2,2}$, represented by  green triangles  numbered $1$ and $2$, to node $2$. Similarly, node $2$ transmits  IVs $v_{3,3}$, $v_{3,4}$ and $v_{3,5}$, represented by  blue squares  numbered $3$, $4$ and $5$, to node $3$. Finally, after the shuffle phase, each node uses the appropriate IVs as input to reduce functions to compute the desired output. {In this example, we see that  Node 1 maps $2$ files, requests 6 IVs, and  the length of each IV is   shorter with $T_1$ bits. Node 2 and 3 each maps 3 files, requests 5 IVs, and the size of each IV is longer with $T_2=T_3$ bits}.


Next, we derive the resulting communication load in this case, denoted by $L^{\rm unicast}(1)$. Note that, the total number of bits in all IVs is $N(T_1+T_2+T_2)$. Also, the number of bits transmitted on the shared-link is $6T_1+5 T_2 + 5 T_3$ as seen in Fig. \ref{fig: uncoded_example} where $6$ IVs of length $T_1$ (red circles), $5$ IVs of length $T_2$ (green triangles), and $5$   IVs of length $T_3$ (blue squares) are transmitted among the nodes. Hence, given $2T_1=T_2=T_3$, we obtain
\begin{align}
L^{\rm unicast}(1) &=
\frac{6T_1+5 T_2 + 5 T_3}{N(T_1+T_2+T_3)} \notag\\
&= \frac{6T_1+10T_1 + 10 T_1}{8(T_1+2T_1+2T_1)} =\frac{13}{20},
\end{align}
which is less than  $L^{\rm LMYA}(1) = \frac{1}{r}\left(1-\frac{r}{K}\right) =  \frac{1}{1}\left(1-\frac{1}{3}\right)  = \frac{2}{3}$  achieved by \cite{li2017fundamental}. This is because the proposed design framework allows varying IV sizes.
\hfill $\triangle$
\end{example}

\begin{example}{\it Coded MapReduce with FLCD}: We study the same network of Example \ref{example: uncoded}, where $K=3$ nodes aim to compute $3$ output functions from $N=8$ input files. Here, nodes $1$, $2$ and $3$ will need to collect all IVs represented by the red circles, green triangles and blue squares, respectively, as shown in Fig. \ref{fig: coded_example}. Unlike the conventional MapReduce, in FLCD, each file is strategically mapped to $r=2$ nodes. Specifically, nodes $1$, $2$ and $3$ map the files of $\mathcal{M}_1=\{ w_1, w_2, w_3, w_6 \}$, $\mathcal{M}_2=\{ w_1, w_3, w_4, w_5, w_7, w_8 \}$ and $\mathcal{M}_3=\{ w_2, w_4, w_5, w_6, w_7, w_8 \}$, respectively. Nodes compute IVs from their locally available files. Similar to before, we let IVs have different lengths, i.e., $2T_1=T_2=T_3$. {In this example, we see that Node 1 maps 4 files, requests 4 IVs, each IV is of a  shorter length $T_1$. Node 2 and 3 each maps 6 files, requests 2 IVs, and each IV is of a longer length $T_2$.}

\begin{figure}[hbt]
  \centering
  \includegraphics[width=2.7in]{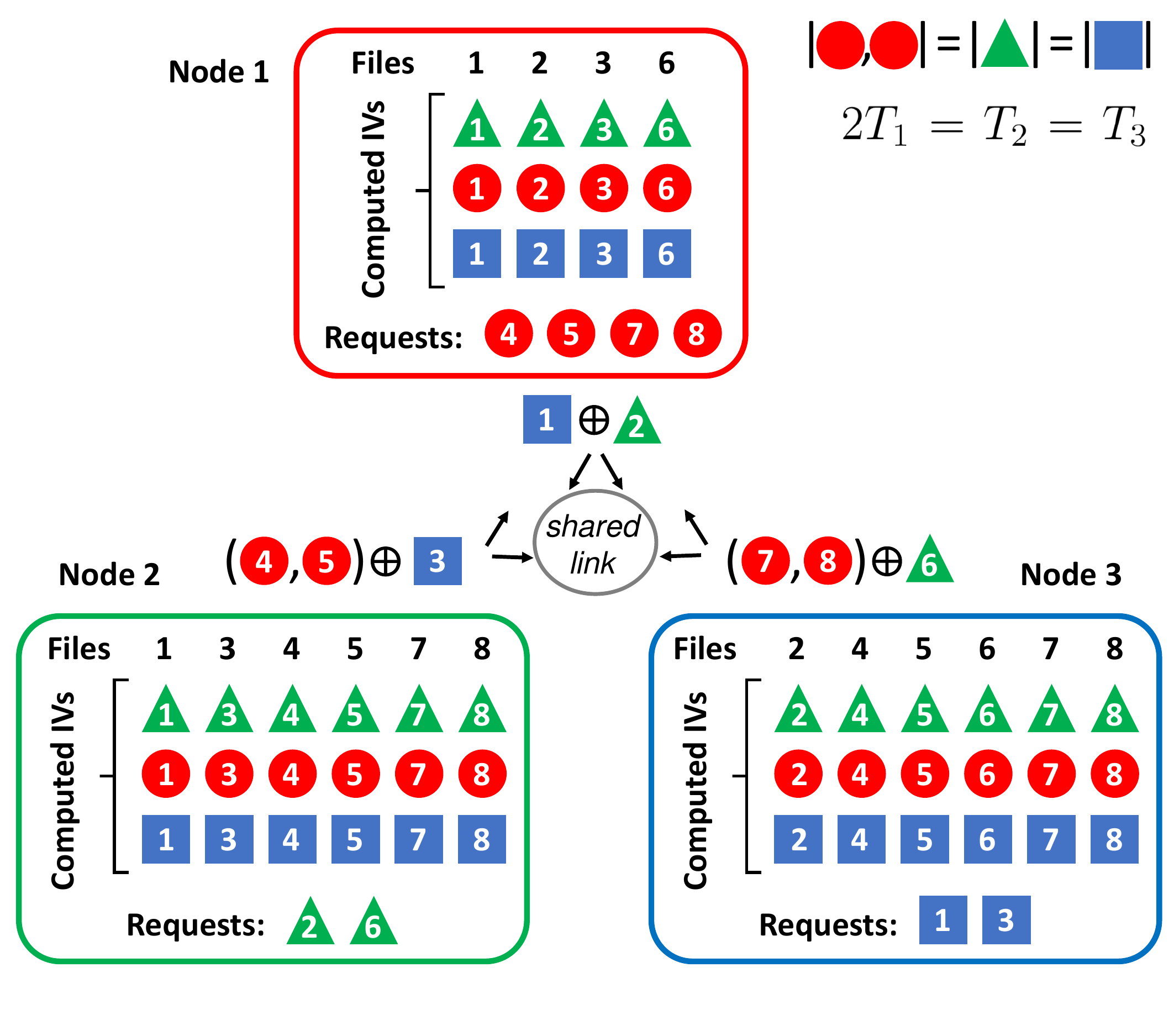}
  \caption{ 
  An example of coded MapReduce using FLCD with $K=3$, $r=2$, and $N=8$. The meaning of the symbols used here are similar to those of Fig. \ref{fig: uncoded_example}. Relative to nodes $2$ and $3$, node 1 requests a greater number of IVs, each with a smaller size of $T_1$ bits. Node 1 requests $4$ IVs, while nodes $2$ and $3$ each request only 2 IVs. Each IV requested by node $2$ or $3$ each have a larger size of $T_3=T_2=2T_1$ bits. Coded multicast messages are transmitted from one node to the other two nodes through a shared link. For instance, node $2$ transmits the XOR of the concatenated message of two red circles $v_{1,4}, v_{1,5}$, intended for node $1$, but available at node $3$, and one blue block $v_{3,3}$, intended for node $3$, but available at node $1$. Both nodes $1$ and $3$ decode their requested IVs using this coded message and locally computes IVs. } \label{fig: coded_example}
\end{figure}

In the shuffle phase, we look for coded multicasting opportunities where a coded message can serve two independent node requests as shown in Fig.~\ref{fig: coded_example}.
In particular, the IVs $v_{3,1}$ and $v_{3,3}$,  represented by blue squares  numbered $1$ and $3$, 
are available at nodes $1$ and $2$ and requested by node $3$.
Similarly, 
 IVs $v_{2,2}$ and $v_{2,6}$, represented by  green squares numbered $2$ and $6$, are available at nodes $1$ and $3$ and requested by node $2$. Hence, node $1$ can transmit the coded pair $v_{3,1}\oplus v_{2,2}$ where ``$\oplus$'' represents the bit-wise XOR operation. 
 Note that $v_{3,1}$ and $v_{2,2}$ have the same size.
 Nodes $2$ and $3$ can recover their requested IVs from this coded multicast using their locally computed IVs.  The transmitted coded message from node $3$ is $(v_{1,7},v_{1,8})\oplus v_{2,6}$. Here, ``$(,)$'' represents the concatenation of two IVs which is necessary since the IVs for output function $1$ are half the size of those for output functions $2$ and $3$. Using locally computed IVs node $1$ can recover $v_{1,7}$ and $v_{1,8}$ and node $2$ can recover $v_{2,6}$. The transmitted coded message from node $2$ can be designed similarly to those of node 3.


As shown from Fig.~\ref{fig: coded_example}, given that $2T_1=T_2=T_3$,  the coded messages transmitted for all nodes have the same size of $T_2=T_3$ bits. For instance, the coded message from node 2 to node 1 and 3 is generated by doing the XOR of  the concatenation of two IVs of length $T_1$ (for a total length of $2T_1$ bits)  and one IV of length $T_3$. The resulting coded message has a length of $2 T_1=T_3$ bits. Hence, the communication load is given by
\begin{align}
L^{\rm FLCD}(2) &=
\frac{K T_2}{N(T_1+T_2+T_3)} \notag\\
&= \frac{3 \cdot 2T_1}{8(T_1+2T_1+2T_1)} = \frac{3}{20},
\end{align}
which is significantly less than that of that of Example 4 where $L^{\rm unicast}(1)=\frac{13}{20}$ due to the use of coded multicasting. It also improves upon the fundamental communication-computation tradeoff  $L^{\rm LMYA}(2) =\frac{1}{6}$, calculated from (\ref{eq:fund_limit}).
This is again because that the proposed flexible design allows varying IV sizes.
\hfill $\triangle$
\end{example}

\section{Achievable communication load and complexity  of FLCD} 
\label{sec: Main Results}

In this section, we will summarize the achievable communication load of the proposed  FLCD and provide a  theoretical complexity comparison against other state-of-the-art designs. Detailed descriptions of the FLCD schemes and empirical evaluations are deferred until Section \ref{sec: FLCD} and Section \ref{sec: empirical evaluation}, respectively. Next, we will first discuss results of the FLCD scheme for the special case of  $K=3$, and  then discuss results for the general FLCD scheme of $K>3$.\footnote{The case of $K \leq 2$ is straightforward. Hence, we do not consider this case.}.

\subsection{Results of FLCD Scheme for  $K=3$}
\label{sec: K3}
When $K=3$, the only non-trivial case is $r=2$.
\footnote{When $K=3$, FLCD is designed only for $r=2$ since the case of $r=1$ is conventional uncoded MapReduce.  When $r=3$, each node maps the entire library and strategic map and shuffle designs are unnecessary.} 
In this case, for FLCD each multicast from any node serve $r=2$ independent node requests. This case allows for arbitrary IV sizes and shows the fundamental tradeoff of \cite{li2017fundamental} does not apply under the more general design framework with heterogeneous IV sizes.

\begin{proposition}\label{prop: FLCD}
   When $K=3$ and $r=2$, { for general IV sizes $T_1,T_2,T_3>0$,} the communication load of FLCD is
  \begin{align}
  \label{eq: L FLCD K3}
  L^{\rm FLCD}(2) = \frac{3T_1T_2T_3}{2(T_1T_2+T_1T_3+T_2T_3)(T_1+T_2+T_3)},
  \end{align}
  where $T_k, k \in [3]$ are the sizes of the IVs for function $k$. The required number of input files is {${\rm LCM}(T_1,T_2,T_3)\times\left(\frac{1}{T_1}+\frac{1}{T_2}+\frac{1}{T_3} \right)$} and the required number of shuffle groups is $1$. 



\end{proposition}
\begin{IEEEproof}
Proposition ~\ref{prop: FLCD} is proved in Appendix~\ref{sec:proof_prop1}.
\end{IEEEproof}



\begin{remark}
When $K=3$, $r=2$ and $T_1=T_2=T_3$, FLCD is equivalent to the LMYA design and $L^{\rm FLCD}(2)=\frac{1}{6}=L^{\rm LMYA}(2)$.  When $T_1$, $T_2$ and $T_3$ are not equal, we have the following corollary.
\begin{corollary}
\label{corollary: 1}
When $K=3$, $r=2$ and $T_1$, $T_2$ and $T_3$ are not equal,
\be
L^{\rm FLCD}(2) <\frac{1}{6}=L^{\rm LMYA}(2).
\ee
\end{corollary}
\begin{IEEEproof}
Corollary~\ref{corollary: 1} is proved in Appendix \ref{sec:proof_corollary1}.
\end{IEEEproof}
From Corollary~\ref{corollary: 1}, it can be seen that in this case, the fundamental limit of \cite{li2017fundamental} is no longer optimal when we allow different IV sizes.
\end{remark}



\subsection{Results of General FLCD Scheme of $K>3$}
 When $K>3$, FLCD achieves a multiplicative communication-computation load tradeoff where each multicast serves $r-1$ nodes. By the design of specific relative IV sizes, FLCD for $K>3$ is flexible in that it operates for any integer $r$ such that $2\leq r\leq \frac{K}{2}$. The performance of FLCD in terms of the communication-computation load tradeoff and the required number of input files and shuffle groups is presented in Theorem \ref{thm: FLCD} in the following.

\begin{theorem}\label{thm: FLCD}
   When $K > 3$ and $2\leq r\leq \frac{K}{2}$, let $m\triangleq\frac{K}{r}$ and $\hat{m}\triangleq \lfloor m \rfloor + 1$,  the communication load of FLCD is
  \begin{align}
  \label{eq: L FLCD}
  L^{\rm FLCD}(r) = \frac{1}{r-1}\left(\frac{\lfloor m \rfloor^2-\lfloor m \rfloor}{\lfloor m \rfloor\hat{m}-m}\right),
  \end{align} and the required number of input files and shuffle groups is
$N=G=\lfloor m \rfloor ^{\left(\hat{m} r - K\right)} \times \hat{m} ^ {\left(K - \lfloor m \rfloor r\right)}$ { and IV sizes are either equal to $T_1'$ or $T_2'$ where $  \lfloor m \rfloor T'_1= (\lfloor m \rfloor - 1)T'_2$.}


\end{theorem}

\begin{IEEEproof}
Theorem~\ref{prop: FLCD} is proved in Appendx~\ref{sec: proof thm 1}.
\end{IEEEproof}

\begin{remark}
For $K>3$, when $\frac{K}{r}=m$ is an integer, we find $L^{\rm FLCD}(r) =L^{\rm KR}(r) = \frac{1}{r-1}\left(1 - \frac{r}{K}\right)$, i.e., the FLCD and the KR designs have the same communication-computation load tradeoff. When $m$ is not an integer, the FLCD can still operate as shown in Section \ref{sec: FLCD}, but the KR scheme is no longer feasible. {Note that while the KR scheme can be used in conjunction with a memory sharing approach to operate on a non-integer $m$,  this will result in a communication load that is greater than the original 
$L^{\rm KR}(r)$ given in (\ref{eq: L_KR}). In contrast, the proposed FLCD is directly designed to  operate on a non-integer $m$ without the need for memory sharing.}  Surprisingly, for an non-integer $m$, we find that when allowing varying IV sizes, 
the communication load of FLCD is less than that of a system with constant IV sizes.  This is described in the following corollary. 
\begin{corollary}
\label{corollary: 2}
When $K>3$ and $m=\frac{K}{r}$ is not an integer, then
\be
L^{\rm FLCD}(r) < \frac{1}{r-1}\left(1 - \frac{r}{K}\right),
\ee
where $r \geq 2$ and $r \in \mathbb{Z}^+$.
\end{corollary}
\begin{IEEEproof}
Corollary~\ref{corollary: 2} is proved in Appendix \ref{sec:proof_cor2}.
\end{IEEEproof}
\end{remark}

\begin{remark}
It can be seen from Section~\ref{sec: FLCD K>3} 
that although the proposed FLCD scheme allows flexible IV lengths, the designed IV lengths and computation loads at each node to achieve (\ref{eq: L FLCD}) will be approximately the same as $m$ becomes large. 
This result is summarized in the following corollary.
\begin{corollary}
\label{corollary: 3}
Assume  $K>3$ and $r \ge 2$. Let $m=\frac{K}{r}$. Then we have the following:
\begin{itemize}
    \item[(i)] Asymptotically equal IV sizes: There exists some $T>0$ such that 
    \be
\lim_{m \rightarrow \infty} T_k = T,  \quad \forall k \in [K].
\label{eq:approx_IV_size}
\ee
\item[(ii)] Asymptotically equal number of files mapped at each node: \be
\lim_{m \rightarrow \infty} |\mathcal{M}_k| = \frac{Nr}{K},  \quad \forall k \in [K]. 
\ee 
\end{itemize}

\end{corollary}

\begin{IEEEproof}
Corollary~\ref{corollary: 3} can be directly obtained in Section~\ref{sec: FLCD K>3}.
\end{IEEEproof}

 From (\ref{eq:approx_IV_size}) of Corollary~\ref{corollary: 3}, we see that when $\frac{K}{r}=m$ is large, the IV sizes of different nodes in the network are approximately equal. In this case,  the constraint of equal IV size commonly used in other designs is relaxed only slightly. { Hence, an important consequence of Corollary~\ref{corollary: 3} is that for small variation in IV size, FLCD can fit a much wider range of parameters since FLCD operates for any integer $r\leq\frac{K}{2}$.} This is opposed to the previous low complexity design \cite{Konstantinidis2020} which only operates for integer $\frac{K}{r}$.
\end{remark}

\subsection{Comparison to State-of-the-Art CDC Designs}

 In Table \ref{table: LNG_points_1}, we list the key parameters $L$ (communication load), $N$(number of files) and $G$ (number of shuffle groups) of different CDC designs considered in this paper.   First, 
as discussed before, the LMYA design \cite{li2017fundamental} has a relatively high complexity. For example, it  requires $N>10^4$  and $G>10^5$  for $K=25$ and $r=5$. As shown in the empirical evaluations (see Section~\ref{sec: empirical evaluation}), a large $G$ greatly negates the promised gain of CDC. Second, the KR design \cite{Konstantinidis2020} has the least complexity in terms of $N$ and $G$, but is rather limited to network parameters with integer $m$. Third, the proposed FLCD scheme can operate on all $(K,r)$ pairs of Table \ref{table: LNG_points_1} and has over $10\times$ reduction in $G$ compared to LMYA \cite{li2017fundamental}. Empirical evaluations of these designs on Amazon EC2 (see Section~\ref{sec: empirical evaluation}) will confirm that FLCD outperforms both of the 
LMYA and KR designs in MapReduce total execution times.

\begin{table*}[t]
\small
  \caption{Flexibility and Complexity of Achievable CDC Designs}
  \label{table: LNG_points_1}
  \centering
  \renewcommand{\arraystretch}{1}
  \begin{tabular}{|ccc|ccc|ccc|ccc|}
    \hline
    \multicolumn{3}{|c}{} &   \multicolumn{3}{c}{LMYA Design \cite{li2017fundamental}}    &   \multicolumn{3}{c}{KR Design \cite{Konstantinidis2020}}   &   \multicolumn{3}{c|}{FLCD}         \\
    $K$     & $r$     & $m$ & $L$ & $N$ & $G$  & $L$ & $N$ & $G$ & $L$ & $N$ & $G$\\
    \hline
    $16$ & $3$  & $5.33$  & $0.27$ & $560$  & $1820$  & $-$ & $-$ & $-$ & $0.41$ & $150$ & $150$   \\
    $16$ & $4$  & $4$     & $0.19$ & $1820$ & $4368$  & $0.25$ & $64$ & $192$ & $0.25$ & $256$ & $256$       \\
    $16$ & $5$  & $3.2$   & $0.14$ & $4368$ & $8008$  & $-$ & $-$ & $-$ & $0.17$ & $324$ & $324$\\
    \hline
    $22$ & $3$  & $4.33$  & $0.29$ & $1540$  & $7315$  & $-$ & $-$ & $-$ & $0.43$ & $392$ & $392$   \\
    $22$ & $4$  & $5.5$     & $0.20$ & $7315$ & $26334$  & $-$ & $-$ & $-$ & $0.27$ & $900$ & $900$       \\
    $22$ & $5$  & $4.4$   & $0.15$ & $26334$ & $74613$  & $-$ & $-$ & $-$ & $0.19$ & $1600$ & $1600$\\
    \hline
    $25$ & $3$  & $8.33$  & $0.29$ & $2300$  & $12650$  & $-$ & $-$ & $-$ & $0.44$ & $576$ & $576$   \\
    $25$ & $4$  & $6.25$     & $0.21$ & $12650$ & $53130$  & $-$ & $-$ & $-$ & $0.28$ & $1512$ & $1512$       \\
    $25$ & $5$  & $5$   & $0.16$ & $53130$ & $177100$  & $0.2$ & $625$ & $2500$ & $0.2$ & $3125$ & $3125$\\
    \bottomrule
  \end{tabular}
\end{table*}


Next, in Section~\ref{sec: FLCD}, we will introduce the general FLCD schemes that achieve  (\ref{eq: L FLCD K3}) and (\ref{eq: L FLCD}), respectively.  

\if{0}
\section{Examples of FLCD for $K=3$ and $r=2$} 
\label{sec: examples}

In this section, we present Examples 4 and 5 to illustrate the key idea of the proposed CDC designs 
by allowing variable IV sizes.  These two examples demonstrate that under our more general design framework by allowing different IV sizes, the fundamental tradeoff, $L^{\rm LMYA}$ of \cite{li2017fundamental} originally derived for the homogeneous IV sizes no longer holds.
Example 4 outlines the conventional uncoded MapReduce approach based on unicast where each input file is mapped at exactly $r=1$ node.  Even for this uncoded case, we show that allowing variable IV sizes results in a lower communication load than that of \cite{li2017fundamental}. In Example 5, FLCD is used for a network of $K=3$ nodes and each input file is mapped to $r=2$ nodes. This example uses coded multicasting, and our design with varying IV sizes again improves on the communication load of \cite{li2017fundamental}.



 \begin{figure}
  \centering
  \includegraphics[width=3in]{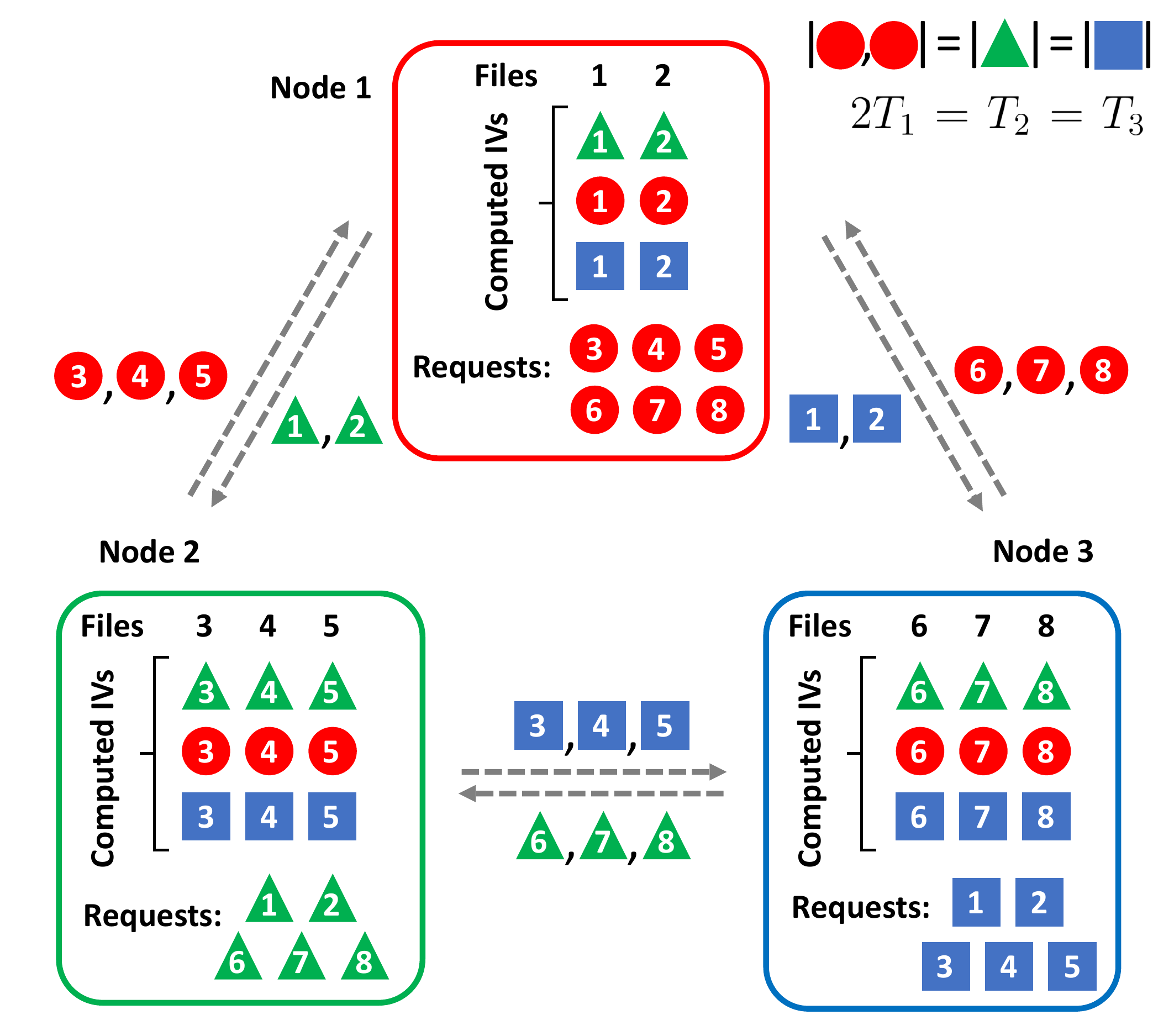} 
  \vspace{-0.4cm}
  \caption{An example of conventional uncoded MapReduce with a computation load of $r=1$. }\label{fig: uncoded_example}
\end{figure}


\begin{example}\label{example: uncoded}{\it Conventional Uncoded MapReduce}: As shown in  Fig.~\ref{fig: uncoded_example}, a network of $K=3$ nodes aims to compute $3$ output functions,  one  assigned to each node. There are $N=8$ input files and each is mapped to  $r=1$ computing node.   Nodes $1$, $2$ and $3$ map the files of $\mathcal{M}_1=\{ w_1, w_2 \}$, $\mathcal{M}_2=\{ w_3, w_4, w_5 \}$ and $\mathcal{M}_3=\{ w_6, w_7, w_8 \}$, respectively. In the map phase, each node computes $3$ IVs, one for each output function, from each of its locally available files. Moreover, the map functions are designed such that $2T_1=T_2=T_3$ and  IVs for node $1$'s reduce function (green triangles) contain half the number of bits of IVs for node $2$
 and $3$'s reduce functions (red circles and blue squares, respectively).

The shuffle phase is necessary so each computing node can collect the needed (or requested) IVs for its reduce function corresponding to its assigned output function. Nodes $1$, $2$ and $3$ will be required to have the access to all the $8$ IVs represented by red circles, green triangles, and blue squares, respectively, as shown in Fig. \ref{fig: uncoded_example}. In order to accomplish this, each node transmits the required IVs to the other two nodes on the shared-link. For instance, node $1$ transmits  IVs $v_{2,1}$ and $v_{2,2}$, represented by  green triangles  numbered $1$ and $2$, to node $2$. Similarly, node $2$ transmits  IVs $v_{3,3}$, $v_{3,4}$ and $v_{3,5}$, represented by  blue squares  numbered $3$, $4$ and $5$, to node $3$. Finally, after the shuffle phase, each node uses the appropriate IVs as input to reduce functions to compute the desired output.


Next, we derive the resulting communication load in this case, denoted by $L^{\rm unicast}(1)$. Note that, the total number of bits in all IVs is $N(T_1+T_2+T_2)$. Also, the number of bits transmitted on the shared-link is $6T_1+5 T_2 + 5 T_3$ as seen in Fig. \ref{fig: uncoded_example} where $6$ IVs of length $T_1$ (red circles), $5$ IVs of length $T_2$ (green triangles), and $5$   IVs of length $T_3$ (blue squares) are transmitted among the nodes. Hence, given $2T_1=T_2=T_3$, we obtain
\be
L^{\rm unicast}(1)=
\frac{6T_1+5 T_2 + 5 T_3}{N(T_1+T_2+T_3)} = \frac{6T_1+10T_1 + 10 T_1}{8(T_1+2T_1+2T_1)} =\frac{13}{20},
\ee
which is less than  $L^{\rm LMYA}(1) = \frac{1}{r}\left(1-\frac{r}{K}\right) =  \frac{1}{1}\left(1-\frac{1}{3}\right)  = \frac{2}{3}$  achieved by \cite{li2017fundamental}. This is because the proposed design framework allows varying IV sizes.
\hfill $\triangle$
\end{example}

\begin{example}{\it Coded MapReduce with FLCD}: We study the same network of Example \ref{example: uncoded}, where $K=3$ nodes aim to compute $3$ output functions from $N=8$ input files. Here, nodes $1$, $2$ and $3$ will need to collect all IVs represented by the red circles, green triangles and blue squares, respectively, as shown in Fig. \ref{fig: coded_example}. Unlike the conventional MapReduce, in FLCD, each file is strategically mapped to $r=2$ nodes. Specifically, nodes $1$, $2$ and $3$ map the files of $\mathcal{M}_1=\{ w_1, w_2, w_3, w_6 \}$, $\mathcal{M}_2=\{ w_1, w_3, w_4, w_5, w_7, w_8 \}$ and $\mathcal{M}_3=\{ w_2, w_4, w_5, w_6, w_7, w_8 \}$, respectively. Nodes compute IVs from their locally available files. Similar to before, we let IVs have different lengths, i.e., $2T_1=T_2=T_3$.

In the shuffle phase, we look for coded multicasting opportunities where a coded message can serve two independent node requests as shown in Fig.~\ref{fig: coded_example}.
In particular, the IVs $v_{3,1}$ and $v_{3,3}$,  represented by blue squares  numbered $1$ and $3$, 
are available at nodes $1$ and $2$ and requested by node $3$.
Similarly, 
 IVs $v_{2,2}$ and $v_{2,6}$, represented by  green squares numbered $2$ and $6$, are available at nodes $1$ and $3$ and requested by node $2$. Hence, node $1$ can transmit the coded pair $v_{3,1}\oplus v_{2,2}$ where ``$\oplus$'' represents the bit-wise XOR operation. 
 Note that $v_{3,1}$ and $v_{2,2}$ have the same size.
 Nodes $2$ and $3$ can recover their requested IVs from this coded multicast using their locally computed IVs.  The transmitted coded message from node $3$ is $(v_{1,7},v_{1,8})\oplus v_{2,6}$. Here, ``$(,)$'' represents the concatenation of two IVs which is necessary since the IVs for output function $1$ are half the size of those for output functions $2$ and $3$. Using locally computed IVs node $1$ can recover $v_{1,7}$ and $v_{1,8}$ and node $2$ can recover $v_{2,6}$. The transmitted coded message from node $2$ can be designed similarly to those of node 3.


As shown from Fig.~\ref{fig: coded_example}, given that $2T_1=T_2=T_3$,  the coded messages transmitted for all nodes have the same size of $T_2=T_3$ bits. For instance, the coded message from node 2 to node 1 and 3 is generated by doing the XOR of  the concatenation of two IVs of length $T_1$ (for a total length of $2T_1$ bits)  and one IV of length $T_3$. The resulting coded message has a length of $2 T_1=T_3$ bits. Hence, the communication load is
\be
L^{\rm FLCD}(2) =
\frac{K T_2}{N(T_1+T_2+T_3)} = \frac{3 \cdot 2T_1}{8(T_1+2T_1+2T_1)} = \frac{3}{20},
\ee
which is significantly less than that of that of Example 4 where $L^{\rm unicast}(1)=\frac{13}{20}$ due to the use of coded multicasting. It also improves upon the fundamental communication-computation tradeoff  $L^{\rm LMYA}(2) =\frac{1}{6}$, calculated from (\ref{eq:fund_limit}).
This is again because that the proposed flexible design allows varying IV sizes.
\hfill $\triangle$
\end{example}

\begin{figure}
  \centering
  \includegraphics[width=3in]{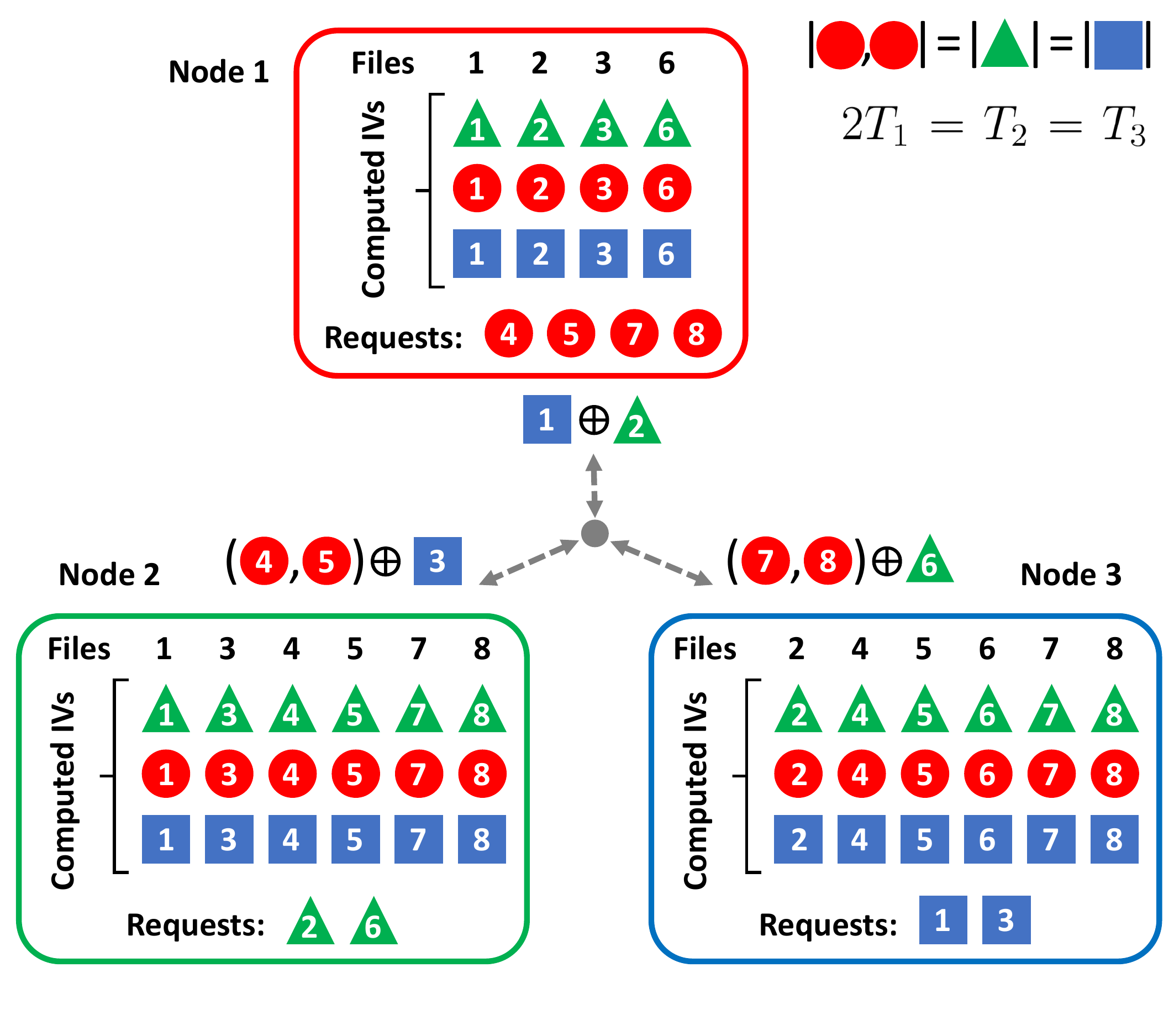}
  \vspace{-0.6cm}
  \caption{An example of coded MapReduce using FLCD with a computation load of $r=2$. } \label{fig: coded_example}
\end{figure}
\fi

\section{Description of the General FLCD Scheme}
\label{sec: FLCD}

In this section, we will first present the general design of FLCD when $K=3$ and $r=2$ and then we will introduce the general design of FLCD for $K>3$.

\subsection{General FLCD Scheme for $K=3$ and $r=2$}
\label{sec: general FLCD when K=3}


{ Given arbitrary IV sizes $T_1,T_2,T_3>0$, we first present the general design of FLCD when $K=3$.} We first split the files into three non-overlapping sets $\mathcal{M}_{\{1,2 \}}$, $\mathcal{M}_{\{1,3 \}}$ and $\mathcal{M}_{\{2,3 \}}$. The files of $\mathcal{M}_{\{1,2 \}}$ are mapped at nodes $1$ and $2$, $\mathcal{M}_{\{1,3 \}}$ are mapped at nodes $1$ and $3$ and $\mathcal{M}_{\{2,3 \}}$ are mapped at nodes $2$ and $3$. Also, given the IV sizes, $T_1$, $T_2$ and $T_3$ bits, the file sets  are defined such that
\begin{align}\label{eq: sile_set_relation}
    |\mathcal{M}_{\{1,2\}}|T_3=|\mathcal{M}_{\{1,3\}}|T_2=|\mathcal{M}_{\{2,3\}}|T_1, 
\end{align}
where $|\mathcal{M}_{\{i,j\}}|$ is the number of files in $\mathcal{M}_{\{i,j\}}$. Then, we define $\mathcal{V}_{\{1,2\}}^3$ as the set of IVs for node $3$'s output function from the files of $\mathcal{M}_{\{1,2\}}$. The IV sets $\mathcal{V}_{\{1,3\}}^2$ and $\mathcal{V}_{\{2,3\}}^1$ are defined similarly. Each IV set $\mathcal{V}_{\{i,j\}}^k$ is split into two equal size sets $\mathcal{V}_{\{i,j\}}^{k,i}$ and $\mathcal{V}_{\{i,j\}}^{k,j}$ to be transmitted in a coded message from node $i$ and $j$, respectively. In the shuffle phase, node $1$ transmits $\mathcal{V}_{\{1,2\}}^{3,1} \oplus \mathcal{V}_{\{1,3\}}^{2,1}$, node $2$ transmits $\mathcal{V}_{\{1,2\}}^{3,2} \oplus \mathcal{V}_{\{2,3\}}^{1,2}$ and node $3$ transmits $\mathcal{V}_{\{1,3\}}^{2,3} \oplus \mathcal{V}_{\{2,3\}}^{1,3}$. Due to (\ref{eq: sile_set_relation}), we can see for each coded transmission, the message sets being XOR'd together have the same length in bits.

Each coded transmission successfully serves independent requests of two nodes simultaneously via the shared-link such that the receiving nodes use locally computed IVs to resolve the requested IVs from this coded transmission. For example, node $2$ receives $\mathcal{V}_{\{1,2\}}^{3,1} \oplus \mathcal{V}_{\{1,3\}}^{2,1}$ from node $1$. Since node $2$ has already computed $\mathcal{V}_{\{1,2\}}^{3,1}$ locally, it can XOR $\mathcal{V}_{\{1,2\}}^{3,1}$ with the received message, $\mathcal{V}_{\{1,2\}}^{3,1} \oplus \mathcal{V}_{\{1,3\}}^{2,1}$,
to recover $\mathcal{V}_{\{1,3\}}^{2,1}$. From the Shuffle phase, each node receives and decodes all needed IVs from files that are not locally available. For example, node $2$ can resolve the IV sets $\mathcal{V}_{\{1,3\}}^{2,1}$ and $\mathcal{V}_{\{1,3\}}^{2,3}$ which collectively contain all IVs from the files $\mathcal{M}_{\{1,3\}}$ that are not available to node $2$ but available at nodes 1 and 3. 
Hence, we can conclude the correctness of FLCD for $K=3$ and $r=2$.


The communication load of FLCD for $K=3$ is shown  in (\ref{eq: L FLCD K3})  and its proof is can be found in Appendix~\ref{sec:proof_prop1}.

\subsection{General FLCD scheme for $K>3$}
\label{sec: FLCD K>3}
Next, we present the proposed FLCD design  
for the general case of  $K>3$.  
It comprises of a strategic file mapping and shuffle design, which centers around supporting varying IV sizes,  such that each node can compute its assigned output function in the reduce phase and the total number of requested IVs bits are kept the same for different nodes. 
\if{0}
\subsection{S3ND}
Assume $K=3$ and $r=2$. For arbitrary non-negative IV sizes, $T_1$, $T_2$ and $T_3$, do the following.
{\bf \em Map Phase}: Define $3$ non-overlapping files sets $\mathcal{M}_{\{1,2 \}}$, $\mathcal{M}_{\{1,3 \}}$ and $\mathcal{M}_{\{2,3 \}}$ where the number of files in each set is defined by
\begin{align}
    \left|\mathcal{M}_{\{i,j\}}\right|=\frac{T_iT_j}{T_1T_2+T_1T_3+T_2T_3}\cdot N.
\end{align}
Then, the files of $\mathcal{M}_{\{1,2 \}}$ are mapped at nodes $1$ and $2$, the files of $\mathcal{M}_{\{1,3 \}}$ are mapped at nodes $1$ and $3$ and the files of $\mathcal{M}_{\{2,3 \}}$ are mapped at nodes $2$ and $3$.
{\bf \em Shuffle Phase}: Define $\mathcal{V}_{\{i,j\}}^k$ as the IVs requested by node $k$ from the files of $\mathcal{M}_{\{i,j \}}$. Split each IV set $\mathcal{V}_{\{i,j\}}^k$ into two IV sets $\mathcal{V}_{\{i,j\}}^{k,i}$ and $\mathcal{V}_{\{i,j\}}^{k,j}$ to be transmitted in a coded message by nodes $i$ and $j$, respectively. Then, nodes $1$ transmits $\mathcal{V}_{\{1,2\}}^{3,1}\oplus\mathcal{V}_{\{1,3\}}^{2,1}$, node $2$ transmits $\mathcal{V}_{\{1,2\}}^{3,2}\oplus\mathcal{V}_{\{2,3\}}^{1,2}$ and $3$ transmits $\mathcal{V}_{\{1,3\}}^{2,3}\oplus\mathcal{V}_{\{2,3\}}^{1,3}$.
\fi
Compared to prior work of \cite{woolsey2020new,woolsey2020cascaded}, which focus on CDC networks with heterogeneous function assignments, 
the FLCD proposed here takes a different approach to explore heterogeneous IV sizes instead of function assignments, assuming that only one reduced function is assigned to each node. This approach leads to a new class of asymptotic homogeneous CDC design proposed here that are amenable for practical implementations due to the reduced packetization. 

Assume $2\leq r\leq\frac{K}{2}$ where $r$ and $K$ are positive integers. Let $m=\frac{K}{r}$ and  $\hat{m} = \lfloor m \rfloor + 1$. We split all the computing nodes into two non-overlapping sets $\mathcal{K}_1$ and $\mathcal{K}_2$. 
Each node in $\mathcal{K}_1$ maps a $\frac{1}{\hat{m}}$ fraction of the entire dataset and each node in $\mathcal{K}_2$ maps a $\frac{1}{\lfloor m \rfloor}$ fraction of the entire dataset.
Moreover, $\mathcal{K}_1$ and $\mathcal{K}_2$ contain $K_1=\hat{m} K - \lfloor m\rfloor\hat{m} r$ and $K_2=\lfloor m\rfloor\hat{m} r - \lfloor m\rfloor K$ nodes, respectively. Note that $K_1+K_2=K$. The number of times that the nodes in $\mathcal{K}_1$ collectively
map the file library is $r_1= K - \lfloor m\rfloor r$. Similarly,
the nodes in $\mathcal{K}_2$ collectively map the entire dataset  $r_2=\hat{m} r - K$ times. Note that $r_1+r_2=r$.  We further split  $\mathcal{K}_1$ and $\mathcal{K}_2$ into $r_1$ and $r_2$, respectively, equally sized non-overlapping sets $\mathcal{K}_1^1,\ldots,\mathcal{K}_1^{r_1}$ and $\mathcal{K}_2^1,\ldots,\mathcal{K}_2^{r_2}$, 
where $|K_1^i|=\frac{K_1}{r_1}=\hat{m}$ and $|K_2^i|=\frac{K_2}{r_2}=\lfloor m\rfloor$. Nodes in each set $\mathcal{K}_\ell^i$ collectively map the file library exactly once. 
Moreover, we design the map functions such that $v_{k,n}$ is of size $T'_1$ bits if $k\in\mathcal{K}_1$ and size $T'_2$ bits if $k\in\mathcal{K}_2$ where $ \lfloor m \rfloor T'_1= (\lfloor m \rfloor - 1)T'_2$. This design choice ensures  each node requests the same number of bits of IVs from each shuffle group. 
When $m$ is large (e.g., $K$ is large and $r$ is fixed), it can be shown that all  IVs will have approximately the same size, i.e,
\be
\lim_{m \rightarrow \infty}  \frac{ T'_1}{T'_2}=  \frac{\lfloor m \rfloor - 1}{ \lfloor m \rfloor} = 1,
\ee
{
and the number of files mapped to each node are approximately the same
\be
\lim_{m \rightarrow \infty}  \frac{ |\mathcal{M}_{k_1}|}{|\mathcal{M}_{k_2}|}=  \frac{N}{\lfloor m\rfloor+1}\cdot\frac{\lfloor m \rfloor}{ N} = 1,
\ee
for any nodes $k_1\in\mathcal{K}_1$ and $k_2\in\mathcal{K}_2$.} This proves Corollary~\ref{corollary: 3}.


{\bf \em Map Phase}: We split the dataset into $N=\hat{m} ^ {r_1}\times \lfloor m \rfloor ^ {r_2}$ files and define $N$ groups, $\mathcal{S}_1,\ldots,\mathcal{S}_N$. Each such group is called a placement group. The placement groups $\mathcal{S}_n, n \in [N]$
consist of all possible sets with cardinality of $r$ nodes such that each set contains exactly one node from every node set $\mathcal{K}_1^1,\ldots,\mathcal{K}_1^{r_1},\mathcal{K}_2^1,\ldots,\mathcal{K}_2^{r_2}$. Each file, $w_n, n \in [N]$ is then placed into every node in $\mathcal{S}_n$. In this way, the library is mapped exactly $r$ times and each node in $\mathcal{S}_n$ maps  file $w_n$ to compute the corresponding  IVs $v_{1,n},\ldots,v_{K,n}$.


{\bf \em Shuffle Phase}: In FLCD, each placement group $\mathcal{S}_n$ also forms a shuffle group. The nodes in $\mathcal{S}_n$ shuffle IVs requested by one node and locally computed by the other $r-1$ nodes in $\mathcal{S}_n$. We define the set of IVs, $\mathcal{V}_n^k$, to be those requested by node $k$ and locally computed by the other nodes in $\mathcal{S}_n$. Then we split $\mathcal{V}_n^k$ into $r-1$ equal size subsets $\mathcal{V}_n^{k,j}$, where $j\in\mathcal{S}_n\setminus k$ and node $j$ is responsible for transmitting the IVs of $\mathcal{V}_n^{k,j}$. Specifically, each node $j\in\mathcal{S}_n$ broadcasts the coded message
    $\bigoplus_{k\in\mathcal{S}_n\setminus j} \mathcal{V}_n^{k,j}$
to the rest of nodes in $\mathcal{S}_n$. It can be seen that due to the requirement that $ \lfloor m \rfloor T'_1= (\lfloor m \rfloor - 1)T'_2$, 
the transmitted messages from node $j$, $\mathcal{V}_n^{k,j}$, $k \in \mathcal{S}_n\setminus j$, have the same length in bits.

The communication load of FLCD for $K>3$ is shown  in (\ref{eq: L FLCD}). The correctness of FLCD for $K>3$ and the proof of (\ref{eq: L FLCD}) can be found in Appendix~\ref{sec: proof thm 1}.

\begin{remark}
When $K>3,$ the size of the IVs in FLCD are exactly the same for integer $m$.  
In this case, we have $K_1=0$, $K_2=K$. This means that all the computing nodes are in $\mathcal{K}_2$ and each maps a $\frac{1}{m}$ fraction of the file library. Hence, in this case, all the IVs are of size $T'_2$ bits.
\end{remark}


\subsection{An Example of FLCD for $K>3$}
\begin{example}
Our goal is to use the FLCD on a network of $K=18$ computing nodes with a computation load of $r=4$. We find $\frac{K}{r}=m=\frac{9}{2}$ is not an integer and the KR design cannot be used. However, by allowing varying IV sizes in the network we can use FLCD. Define $\hat{m}=\lfloor m \rfloor+1=5$, we split the nodes into two sets $\mathcal{K}_1$ and $\mathcal{K}_2$ of size $K_1=\hat{m} K - \lfloor m\rfloor\hat{m} r = 10$ and $K_2=\lfloor m\rfloor\hat{m} r - \lfloor m\rfloor K = 8$ nodes, respectively. In particular, the $10$ nodes of $\mathcal{K}_1$ will each map $\frac{1}{\hat{m}}=\frac{1}{5}$ of the files and the $8$ nodes of $\mathcal{K}_2$ will each map $\frac{1}{\lfloor m \rfloor}=\frac{1}{4}$ of the files. These node sets are each split into $2$ equally sized disjoint subsets such that  $\mathcal{K}_1=\mathcal{K}_1^1\cup\mathcal{K}_1^2$ and $\mathcal{K}_2=\mathcal{K}_2^1\cup\mathcal{K}_2^2$. The file library is split into $N=5^2\cdot4^2=400$ equally sized files where a file is mapped at a set of $r=4$ nodes, $\mathcal{S}_n$ with one node from each set of  $\{\mathcal{K}_1^1,\mathcal{K}_1^2,\mathcal{K}_2^1,\mathcal{K}_2^2\}$. As an example, let $\mathcal{S}_1=\{1,2,3,4\}$ where nodes $1$, $2$, $3$ and $4$ belong to the sets $\mathcal{K}_1^1$, $\mathcal{K}_1^2$, $\mathcal{K}_2^1$, and $\mathcal{K}_2^2$, respectively. A file is mapped to these nodes that is not mapped to any other of the $14$ nodes.

Each set $\mathcal{S}_n$, $n\in[N]$ also represents a shuffle group. Fig. \ref{fig: Kgt3_example} shows the IVs requested and transmitted by the nodes of $\mathcal{S}_1=\{1,2,3,4\}$. Each node requests IVs that are locally computed at the other nodes, presenting multicast opportunities. For example, the nodes of $\mathcal{S}_1\setminus \{1\} = \{ 2,3,4 \}$ also form placement groups with the $4$ nodes of $\mathcal{K}_1^1\setminus \{ 1 \}$. Therefore, there are $4$ files available to the nodes of $\{2,3,4\}$, but not node $1$. Without loss of generality, let these files be $w_1$, $w_2$, $w_3$ and $w_4$, then node $1$ requests the IVs of $\mathcal{V}_1^1=\{ v_{1,1},v_{1,2},v_{1,3},v_{1,4}\}$ from the shuffle group $\mathcal{S}_1$. Similarly, node $2$ requests the $4$ IVs of $\mathcal{V}_1^2=\{v_{2,5},v_{2,6},v_{2,7},v_{2,8}\}$. Then, we see nodes $3$ and $4$ each request $3$ IVs from $\mathcal{S}_1$ because the nodes of $\{1,2,4\}$ and $\{1,2,3 \}$ form placement groups with the $3$ nodes of $\mathcal{K}_2^1\setminus\{ 3\}$ and $\mathcal{K}_2^2\setminus\{ 4\}$, respectively. Again, without loss of generality, node $3$ requests the IVs of $\mathcal{V}_1^3=\{v_{3,9},v_{3,10},v_{3,11}\}$ and node $4$ requests the IVs of $\mathcal{V}_1^4=\{v_{4,12},v_{4,13},v_{4,14}\}$ from $\mathcal{S}_1$. Since the IVs requested by nodes $1$ and $2$ are $T_1'$ bits each and the IVs requested by nodes $3$ and $4$ are $T_2'$ bits each, we see each node requests the same number of bits from this shuffle group $\mathcal{S}_1$ since $\lfloor m \rfloor T_1'=4T_1'=3T_2'=(\lfloor m \rfloor-1)T_2'$.

\begin{figure}
  \centering
  \includegraphics[width=3in]{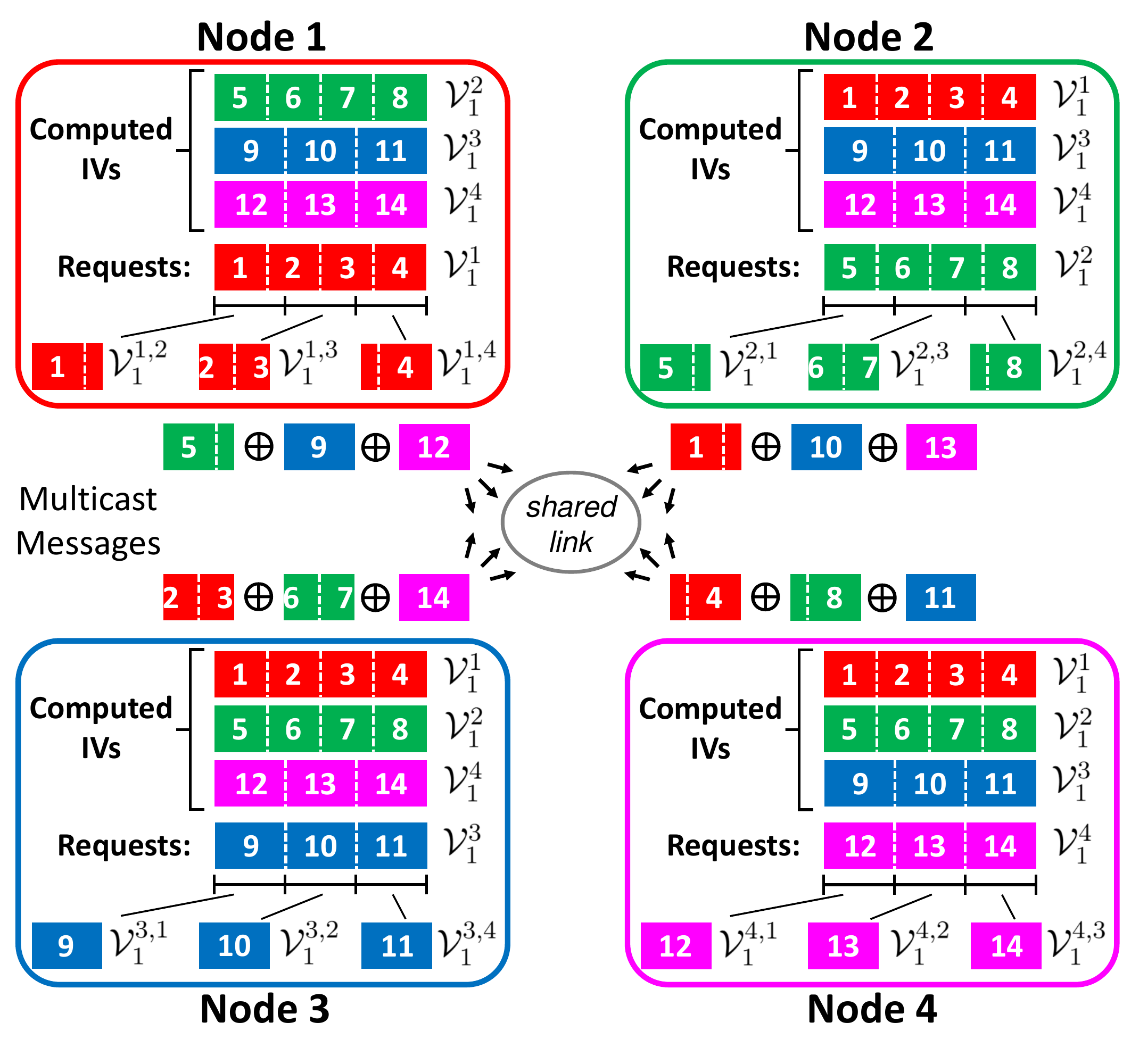}
  \caption{
  Illustration of the data shuffle within a specific shuffle group $\mathcal{S}_1$ of the FLCD scheme for a CDC network with  $K=18$ and $r=4$. Symbols are defined similarly to those of Fig. 2, with the addition of reduce function $h_4$, assigned to node $4$, whose IVs are shown as magenta rectangles.
  Note that only the IVs of interests within $\mathcal{S}_1$, i.e., those requested by one node of $\mathcal{S}_1$ and locally computed by the other $3$ nodes, are shown. Here, nodes $1$ and $2$ each requests $4$ IVs, each of a shorter length $T_1'$ bits; nodes $3$ and $4$ each requests 3 IVs,  each  of a longer length $T_2'=\frac{4}{3}T_1'$ bits. As a result, the total number of requested bits from each  node is the same, shown as equal width of   $\mathcal{V}_1^{1}$, $\mathcal{V}_1^{2}$, $\mathcal{V}_1^{3}$, $\mathcal{V}_1^{4}$ for each node. 
  The requested IVs of each node are concatenated and split into $3$ messages. Node $i$ transmits a coded message of $3$ XOR'd IVs (or fractions of IVs), each intended for a  node in $\mathcal{S}_1\setminus \{i\} $. Given the IVs computed from locally available files, each node can recover its requested IVs from the coded transmissions. 
  } \label{fig: Kgt3_example}
\end{figure}

Fig. \ref{fig: Kgt3_example} depicts the IVs of $\mathcal{V}_1^1$ (red squares), $\mathcal{V}_1^2$ (green squares), $\mathcal{V}_1^3$ (blue rectangles), and $\mathcal{V}_1^4$ (magenta rectangles). In particular, the width of the IVs reflect their relative size. In practice, the IVs requested by a particular node will be concatenated as shown in Fig.~\ref{fig: Kgt3_example} where the IVs are lined up side-by-side. We visualize that each node requests the same amount because the width of the concatenated messages are the same. Then, each concatenated IV set is split into $r-1=3$ messages to be transmitted by $3$ different nodes. For example,  $\mathcal{V}_1^1$ is split into $\mathcal{V}_1^{1,2}$,  $\mathcal{V}_1^{1,3}$, and $\mathcal{V}_1^{1,4}$ to be transmitted by nodes $2$, $3$ and $4$, respectively. Note that, in practice, $\mathcal{V}_1^{1,i}$, $i\in\{2,3,4 \}$ each contain fractions of IVs and not necessarily whole IVs in order to split $\mathcal{V}_1^1$ into $3$ equal size subsets.

Each node $i\in\mathcal{S}_1=\{1,2,3,4\}$ transmits $\bigoplus_{j\neq i}\mathcal{V}_1^{j,i}$ to the other nodes of $\mathcal{S}_1$. For example, node $1$ transmits the coded combination of $\mathcal{V}_1^{2,1}$ (green rectangle that includes $v_{2,5}$ and a fraction of $v_{2,6}$), $\mathcal{V}_1^{3,1}=\{v_{3,9}\}$ (blue rectangle with the number $9$) and $\mathcal{V}_1^{4,1}$ (magenta rectangle with the number $12$). The size of the transmission from each node is $\frac{4}{3}T_1'=T_2'$ bits. Accounting for each shuffle group $\mathcal{S}_n$, $n\in[N]$, the communication load is $L^{\rm FLCD}=\frac{400 \cdot 4 \cdot T_2'}{N ( |\mathcal{K}_1|\cdot T_1'+|\mathcal{K}_2|\cdot T_2')}\approx0.2581$ where we normalize by the total bits of all IVs which is $N ( |\mathcal{K}_1|\cdot T_1'+|\mathcal{K}_2|\cdot T_2')$ bits.

\end{example}

\if
{\it Communication Load}: We count the number of bits transmitted. By design, the number of bits in each IV set $\mathcal{V}_n^{k,j}$ is $\lfloor m \rfloor T'_1$ bits for the following reason. If $k\in\mathcal{K}_1^i\subseteq\mathcal{K}_1$, there are  $|\mathcal{K}_1^i|-1=\hat{m}-1=\lfloor m \rfloor$ files that the nodes of $\mathcal{S}_n\setminus k$ have access to but node $k$ does not. These files are defined by the files mapped to the nodes $\mathcal{S}_n\setminus k$ and a node $k'\in\mathcal{K}_1^i\setminus k$.  Therefore node $k$ requests $\lfloor m \rfloor$ IVs each of size $T'_1$ bits from the nodes of $\mathcal{S}_n\setminus k$. Similarly, if $k\in\mathcal{K}_2^i\subseteq\mathcal{K}_2$, there are $|\mathcal{K}_2^i|-1=\lfloor m \rfloor - 1$ IVs each of size $T_2'$ bits that node $k$ requests from the nodes of $\mathcal{S}_n\setminus k$. Since $ \lfloor m \rfloor T'_1= (\lfloor m \rfloor - 1)T'_2$, each IV set $\mathcal{V}_n^{k,j}$ is $\lfloor m \rfloor T'_1$ bits. Consider all $N$ node groups $\mathcal{S}_n$ for which $r$ nodes each send a message of size $\frac{\lfloor m \rfloor T'_1}{r-1}$ bits, the communication load is
\begin{align}
    L &= \frac{1}{N(K_1T'_1+K_2T'_2)} \cdot N \cdot r \cdot \frac{\lfloor m \rfloor T'_1}{r-1} \\
    &= \frac{1}{r-1}\cdot \frac{r\lfloor m \rfloor(\lfloor m \rfloor - 1)}{(\hat{m}K-\lfloor m \rfloor\hat{m}r)(\lfloor m \rfloor - 1)+(\lfloor m \rfloor \hat{m} r -\lfloor m \rfloor K)\lfloor m \rfloor} \\
    &=\frac{1}{r-1}\left( \frac{\lfloor m \rfloor^2-\lfloor m \rfloor}{\lfloor m \rfloor \hat{m} - m}\right)
\end{align}
\fi

\section{Empirical Evaluation on Amazon EC2}
\label{sec: empirical evaluation}

\subsection{Experiment Setup}

In order to evaluate the effectiveness of the proposed FLCD approach, we perform a TeraSort algorithm \cite{noll2011benchmarking} on Amazon EC2 with $K=16,22,25$ worker nodes and an additional master node. Each computing node is a \verb+t2.large+ EC2 instance with $2$ vCPUs, $8$ GiB of RAM and $24$ GB of solid-state drive (SSD) storage. We developed Python software to implement a TeraSort algorithm using the proposed FLCD,  LMYA \cite{maddah2014fundamental}, KR  \cite{Konstantinidis2020}, and the conventional uncoded design. Nodes sort $12$ GB of data comprised of $6\times 10^8$ key-value pairs (KVs) in total. Each key is a $16$-bit unsigned integer (\verb+uint16+) and each value a length-$9$ array of $16$-bit unsigned integers. Each node is assigned an output function to sort KVs with keys in a specific range. We design the map and reduce functions using the method outlined in Examples \ref{example: TeraSortFunctions} and \ref{example: varyingIVs} so that the length of the IVs satisfy the requirement of FLCD and $  \lfloor m \rfloor T'_1= (\lfloor m \rfloor - 1)T'_2$ {for non-integer $m$.} as well as for the homogeneous requirements of the LMYA and KR designs.  The map functions hash the KVs by placing KVs in bins based on their keys. The bins correspond to the specific range of keys each node is assigned to sort. We use the open Message Passing Interface (MPI) library to facilitate the inter-node communications. To prevent bursty communication rates, the incoming and outgoing traffic rate of each computing node is limited to $100$ Megabits per second (Mbps) using the Linux \verb+tc+ command. The execution is split into $6$ steps described as follows.


\begin{enumerate}
  \item
  {\it CodeGen}: The worker nodes define placement and shuffle groups and reduce function assignments. The placement groups define partitions of the data and the set of KVs that each node will map based on the specific CDC design. The shuffle groups are defined using the  MPI \verb+Create+ function to create a new MPI-communicator and facilitate the shuffle phase.

  \item
  {\it Map}: The worker nodes load data from the solid-state drive (SSD) and use map functions to hash KVs into bins defined by the reduce functions, or the range of values the nodes are responsible for sorting.

  \item
   {\it Encode}: Based on the CDC design, the worker nodes form the coded messages of IVs that will be used for the multicast transmissions. The IVs are combined using bit-wise XOR and concatenation operations. Note that this step does not apply to the corresponding uncoded design.

  \item
  {\it Shuffle}: Nodes sequentially transmit the (coded) messages to the other nodes in the same shuffle groups based on the  shuffle design of the specific CDC design. For data transmission, the coded designs use the MPI \verb+bcast+ function and the uncoded design uses the MPI \verb+scatter+ function.

  \item
  {\it Decode}: Using the received and locally computed coded messages, the nodes resolve the necessary IVs for their assigned reduce functions. Note that this step does not apply to the uncoded design.

  \item
  {\it Reduce}: The nodes execute their assigned reduce functions to sort the IVs within their corresponding assigned range. In this way, the data set is sorted across the computing network.
\end{enumerate}

We provide the developed Python code for this evaluation on the Github page {\url{https://github.com/C3atUofU/Coded-Distributed-Computing-over-AWS}}.

\begin{figure*}
  \centering
  \includegraphics[width=4.4in]{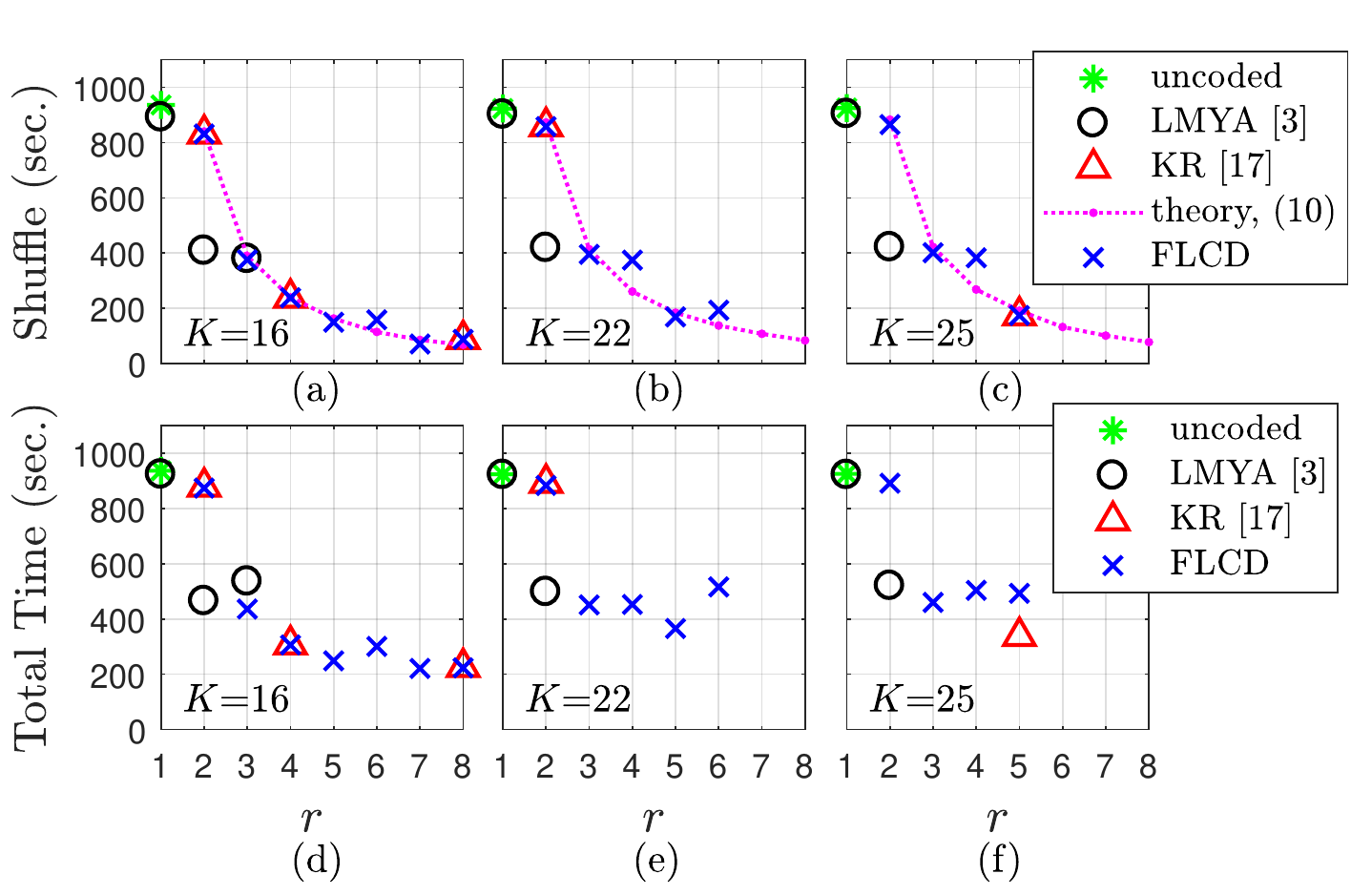} 
  \caption{Empirical evaluations of the proposed FLCD,  LMYA \cite{li2017fundamental}, KR  \cite{Konstantinidis2020}  on Amazon EC2 for implementing the TeraSort Algorithm using $K=16, 22, 25$ computing nodes. In the first row,  (a)-(c) show shuffle time versus computation load $r$ for the three schemes and the theoretical prediction of shuffle time from (\ref{eq: L FLCD}). 
  In the second row, (c)-(d) show total time versus $r$.
  }
  \label{fig: eval}
\end{figure*}
\begin{table*}[h]
\small
  \caption{Empirical Evaluation with $K=16$ worker nodes}
  \label{table: LNG_points 1}
  \centering
  \renewcommand{\arraystretch}{1}
  \begin{tabular}{|c|c|c||c|c|c|c|c|c|c|c|}
    \hline
    Design & $r$ & \shortstack{\\ IV \\ size \\ratio} & \shortstack{\\ CodeGen \\ (sec.)}        & \shortstack{\\ Map \\ (sec.)} & \shortstack{\\ Encode \\ (sec.)} & \shortstack{\\ Shuffle \\ (sec.)} & \shortstack{\\ Decode \\ (sec.)}  & \shortstack{\\ Reduce \\ (sec.)} & \shortstack{\\ Total \\ Time \\ (sec.)} & Speedup \\
    \hline
    Uncoded &$1$&$1$& $0.05$  & $14.94$ & $-$  & $906.46$  & $-$ & $14.63$ & $936.07$   & $-$\\
    \hline
     LMYA \cite{li2017fundamental} &$1$&$1$ & $0.79$  & $15.14$ & $0.81$  & $891.69$  & $0.79$ & $13.55$ & $922.76$   & $1.01\times$\\
     LMYA \cite{li2017fundamental} &$2$&$1$ & $15.23$  & $27.34$ & $1.10$  & $409.37$  & $0.58$ & $11.40$ & $465.01$   & $2.01\times$\\
     LMYA \cite{li2017fundamental}& $3$&$1$ & $101.37$  & $39.72$ & $1.16$  & $379.90$  & $0.62$ & $13.143$ & $535.91$   & $1.75\times$\\
    \hline
     KR \cite{Konstantinidis2020}& $2$&$1$ & $0.437$  & $30.45$ & $0.77$  & $831.83$  & $0.76$ & $13.65$ & $877.89$   & $1.07\times$\\
     KR \cite{Konstantinidis2020}& $4$&$1$ & $1.83$  & $55.32$ & $0.74$  & $238.13$  & $0.44$ & $10.74$ & $307.21$   & $3.05\times$\\
     KR \cite{Konstantinidis2020}& $8$&$1$ & $1.08$  & $122.46$ & $1.13$  & $88.50$  & $0.32$ & $11.81$ & $225.31$   & $4.15\times$\\
    \hline
     FLCD &$2$&$1$ & $0.29$  & $30.85$ & $1.10$  & $831.12$  & $0.63$ & $9.16$ & $873.14$   & $1.07\times$\\
     FLCD &$3$&$4:5$ & $0.89$  & $45.51$ & $1.53$  & $376.31$  & $0.63$ & $10.89$ & $435.75$   & $2.15\times$\\
     FLCD &$4$&$1$ & $2.04$  & $49.09$ & $1.60$  & $238.66$  & $0.49$ & $14.02$ & $305.91$   & $3.06\times$\\
     FLCD &$5$&$2:3$ & $5.08$  & $75.28$ & $1.90$  & $150.34$  & $0.49$ & $15.44$ & $248.52$   & $3.77\times$\\
     FLCD &$6$&$1:2$ & $3.60$  & $109.86$ & $2.22$  & $159.08$  & $0.56$ & $25.32$ & $300.65$   & $3.11\times$\\
     FLCD &$7$&$1:2$ & $3.03$  & $125.84$ & $1.96$  & $71.51$  & $0.46$ & $18.18$ & $220.98$   & $4.24\times$\\
     FLCD &$8$&$1$ & $3.43$  & $115.23$ & $1.89$  & $88.14$  & $0.39$ & $13.89$ & $222.97$   & $4.20\times$\\
    \hline
  \end{tabular}
\end{table*}

\subsection{Results}

Evaluation results are shown in Fig.~\ref{fig: eval} and Table~\ref{table: LNG_points 1} ($K=16$), where  shuffle times for different $K$ are shown in Fig.~\ref{fig: eval}(a) to Fig.~\ref{fig: eval}(c) and total times are shown in Fig.~\ref{fig: eval}(d) to Fig.~\ref{fig: eval}(f). In addition, the ``Speedup" column in Table~\ref{table: LNG_points 1} refers to the factor speed-up compared to conventional uncoded MapReduce. The following observations are made based on these results.
\begin{itemize}
    \item For most points in Fig.~\ref{fig: eval}(a) to Fig.~\ref{fig: eval}(c), the shuffle time decreases proportionally  to  $r$ and
almost
coincide with the theoretical results  (\ref{eq: L FLCD}). This is the first time that theoretical predictions of the shuffle time of a CDC design are validated by empirical evaluations for a large range of $r$. 
{There are a few points in (b) and (c) where the shuffle times lie above  (\ref{eq: L FLCD}), possibly due to the underlying topology of EC2 and the MPI protocol. For instance, the efficiency and overhead of the multicast changes depending on the number of nodes in the multicast group, whereas  this is assumed to be constant in calculating the theoretic prediction  (\ref{eq: L FLCD}).}

\item In Fig.~\ref{fig: eval}(a) to Fig.~\ref{fig: eval}(c), for most of the points,  total time decreases significantly with increasing $r$ despite the time of Map Phase increasing greatly as $r$ grows due to increased computations at each node.
This demonstrates the multiplicative gain of CDC holds even for the total time.
\item The proposed FLCD scheme outperforms LMYA when comparing total time. From Fig.~\ref{fig: eval}(d) to Fig.~\ref{fig: eval}(f), for each value of $K$, with the choice of $r$  that minimizes the total time, FLCD has 
a total time $12\% \sim 52\%$ lower than  LMYA. 
\item  While the FLCD and  KR  have similar shuffle and total time, the FLCD has greater flexibility.  Table~\ref{table: LNG_points 1} shows that when $K=16$, the scenario of $r=5$ cannot be achieved by the KR scheme, and the gain in terms of the total time of FLCD is $19\%$ compared to KR scheme ($r=4$). This observation is important because in practical networks may be storage limited and $r=5$ may be an upper limit for example. {Note that for the case of $K=25$ and $r=5$, the KR scheme has the lowest total time, possibly because the KR scheme requires a much smaller $N$ than that of the FLCD for this setting (see Table 1) and a smaller CodeGen time and Map time (see Table IV).} 
\item From Table~\ref{table: LNG_points 1}, the FLCD ($r=5$) has a $47\%$ reduction in total time compared to the LMYA scheme with $r=2$. Due to the high complexity of the LMYA,  the maximum implementable $r$ is limited to $3$.
\item Table~\ref{table: LNG_points 1} shows a $2.15 \sim 4.24\times$ speed-up of  the FLCD design compared to the conventional uncoded MapReduce approach.
\end{itemize}


Additional evaluation results are also provided in Tables~\ref{table: LNG_points 3} and \ref{table: LNG_points 4}, 
which include a detailed break down of the times of each step for $K=22,25$ worker nodes similar to the case for $K=16$.
These evaluations show similar behavior of all the schemes considered in this paper and demonstrate the significant advantage of the proposed FLCD. For example, in Table~\ref{table: LNG_points 3}, we see that the KR and LMYA scheme are only feasible for $r=2$ and $r=1$, respectively, but the FLCD scheme allows for up to $r=6$.
{In addition, we observe from Tables~\ref{table: LNG_points 3} and \ref{table: LNG_points 4} a clear trend that  the IV ratio  approaches 1 as $\frac{K}{r}$ increases (or equivalently $r$ decreases). This confirms that the proposed design leads to asymptotic homogeneous systems for which the reduced communication load and implementation complexity are achieved with only small variations in IV sizes.}

\begin{table*}[bht]
\small
  \caption{Empirical Evaluation with $K=22$ worker nodes }
  \label{table: LNG_points 3}
  \centering
  \renewcommand{\arraystretch}{1}
  \begin{tabular}{|c|c|c||c|c|c|c|c|c|c|c|}
    \hline
    Design & $r$ &\shortstack{\\ IV \\ size \\ratio} & \shortstack{\\ CodeGen \\ (sec.)}        & \shortstack{\\ Map \\ (sec.)} & \shortstack{\\ Encode \\ (sec.)} & \shortstack{\\ Shuffle \\ (sec.)} & \shortstack{\\ Decode \\ (sec.)}  & \shortstack{\\ Reduce \\ (sec.)} & \shortstack{\\ Total \\ Time \\ (sec.)} & Speedup \\
    \hline
    Uncoded &$1$& $1$ & $0.02$  & $12.20$ & $1$  & $903.78$  & $-$ & $6.96$ & $922.95$   & $-$\\
    \hline
     LMYA \cite{li2017fundamental} &$1$& $1$ & $2.92$  & $7.79$ & $0.61$  & $901.73$  & $0.41$ & $8.56$ & $921.40$   & $1.00\times$\\
     LMYA \cite{li2017fundamental} &$2$& $1$ & $44.00$  & $24.28$ & $0.89$  & $419.50$  & $0.43$ & $8.83$ & $497.93$   & $1.85\times$\\
    \hline
     KR \cite{Konstantinidis2020} &$2$& $1$ & $5.17$  & $16.85$ & $0.58$  & $858.61$  & $0.56$ & $9.71$ & $891.46$   & $1.04\times$\\
    \hline
     FLCD &$2$& $1$ & $0.74$  & $15.82$ & $0.86$  & $857.72$  & $0.51$ & $7.14$ & $882.788$   & $1.05\times$\\
     FLCD &$3$& $6:7$ & $3.26$  & $40.22$ & $1.25$  & $395.67$  & $0.45$ & $9.93$ & $450.76$   & $2.05\times$\\
     FLCD &$4$& $5:6$ & $16.88$  & $48.11$ & $1.60$  & $374.86$  & $0.53$ & $10.49$ & $452.37$   & $2.04\times$\\
     FLCD &$5$& $4:5$ & $107.76$  & $75.53$ & $2.05$  & $169.41$  & $0.68$ & $10.03$ & $365.46$   & $2.53\times$\\
     FLCD &$6$& $3:4$ & $187.30$  & $118.03$ & $3.12$  & $193.69$  & $1.08$ & $12.62$ & $515.83$   & $1.79\times$\\
    \hline
  \end{tabular}
\end{table*}

\begin{table*}[bht]
\small
  \caption{Empirical Evaluation with $K=25$ worker nodes}
  \label{table: LNG_points 4}
  \centering
  \renewcommand{\arraystretch}{1}
  \begin{tabular}{|c|c|c||c|c|c|c|c|c|c|c|}
    \hline
    Design & $r$&\shortstack{\\ IV \\ size \\ratio} & \shortstack{\\ CodeGen \\ (sec.)}        & \shortstack{\\ Map \\ (sec.)} & \shortstack{\\ Encode \\ (sec.)} & \shortstack{\\ Shuffle \\ (sec.)} & \shortstack{\\ Decode \\ (sec.)}  & \shortstack{\\ Reduce \\ (sec.)} & \shortstack{\\ Total \\ Time \\ (sec.)} & Speedup \\
    \hline
    Uncoded &$1$& $1$& $0.06$  & $13.94$ & $-$  & $904.55$  & $-$ & $6.24$ & $924.78$   & $-$\\
    \hline
     LMYA \cite{li2017fundamental} &$1$& $1$ & $3.24$  & $6.79$ & $0.52$  & $903.93$  & $0.37$ & $6.26$ & $921.10$   & $1.00\times$\\
     LMYA \cite{li2017fundamental}& $2$& $1$ & $74.66$  & $13.82$ & $0.82$  & $421.94$  & $0.43$ & $8.47$ & $520.14$   & $1.78\times$\\
    \hline
     KR \cite{Konstantinidis2020}& $5$& $1$ & $125.21$  & $28.37$ & $0.94$  & $174.34$  & $0.56$ & $8.06$ & $337.50$   & $2.74\times$\\
    \hline
     FLCD &$2$& $11:12$ & $1.04$  & $15.13$ & $0.78$  & $865.01$  & $0.39$ & $8.46$ & $890.82$   & $1.04\times$\\
     FLCD &$3$& $7:8$ & $9.99$  & $37.96$ & $1.19$  & $401.59$  & $0.44$ & $8.67$ & $459.83$   & $2.01\times$\\
     FLCD &$4$& $5:6$ & $57.57$  & $51.24$ & $1.52$  & $383.38$  & $0.53$ & $9.16$ & $503.51$   & $1.84\times$\\
     FLCD &$5$& $1$ & $212.71$  & $93.77$ & $2.17$  & $174.78$  & $0.79$ & $8.39$ & $492.62$   & $1.88\times$\\
    \hline
  \end{tabular}
\end{table*}

\section{Conclusions}
\label{sec: conclusions}


In this work, we developed a new flexible, low complexity design (FLCD) to expedite computing platforms such as MapReduce and Spark by trading increased local computation with reduced communication across the network.  Built upon  a combinatorial design for the Map and Shuffle phase, the FLCD schemes  utilize the design freedom in defining map and reduce functions to facilitate varying IV sizes under a general MapReduce framework. This new approach led to an interesting class of asymptotic homogeneous CDC systems that can adapt to a wide range of network parameters and facilitate low complexity implementation, while requiring only small variations in the IV sizes. We provided the most comprehensive empirical evaluations to date on Amazon EC2 for the comparisons of the CDC schemes. Our evaluations of the FLCD covered noticeably more network configurations than previous designs permitted and showed substantial reductions of 12$\%$-52$\%$ in total time under the same network parameters. These successfully validated the flexibility and low complexity of the FLCD schemes. 
{An interesting direction for future work is to explore more communication efficient CDC designs with flexible IV sizes that can serve $r$ nodes within each shuffle group, as opposed to serving only $r-1$ nodes as in the present FLCD design for $K>3$. This has the potential to generalize the proposed FLCD design  for the special case of $K=3$ to arbitrary $K$, and  possibly lead to a better communication-computation trade off in this general MapReduce framework. }

\appendices

\section{The Proof of Proposition \ref{prop: FLCD}}
\label{sec:proof_prop1}
  We consider the FLCD scheme for $K=3$ and $r=2$. It can be seen directly that the FLCD scheme is correct from its description in Section~\ref{sec: general FLCD when K=3}. Here, we will derive the communication load (\ref{eq: L FLCD K3}).

From the FLCD description in Section~\ref{sec: general FLCD when K=3}, it can be seen that this scheme is correct straightforwardly. Note that, by (\ref{eq: sile_set_relation}), the number of total bits of each IV set $\mathcal{V}_{\{i,j\}}^k$ is the same since it contains $|\mathcal{M}_{\{ i,j\}}|$ IVs of size $T_k$ bits each. Let the number of bits in each IV set $\mathcal{V}_{\{i,j\}}^k$ be $B$, then $|\mathcal{M}_{\{ i,j\}}|T_k=B$ and we obtain
\begin{align}
    &L^{\rm FLCD}(2) = \frac{3(B/2)}{N(T_1+T_2+T_3)} \notag 
    \end{align}
\begin{align} 
    &= \frac{3B}{2(|\mathcal{M}_{\{ 1,2\}}|+|\mathcal{M}_{\{ 1,3\}}|+|\mathcal{M}_{\{ 2,3\}}|)(T_1+T_2+T_3)} \notag\\
    &= \frac{3B}{2\left(\frac{B}{T_3}+\frac{B}{T_2} + \frac{B}{T_1}  \right)(T_1+T_2+T_3)} \notag\\
    &=\frac{3T_1T_2T_3}{2(T_1T_2+T_1T_3+T_2T_3)(T_1+T_2+T_3)}.
\end{align}

Hence, we finish the proof of Proposition \ref{prop: FLCD}.

\section{Proof of Corollary~\ref{corollary: 1}}
\label{sec:proof_corollary1}
In this section, we will prove Corollary~\ref{corollary: 1}, which states that $L^{\rm FLCD}(2) <\frac{1}{6}=L^{\rm LMYA}(2)$ when $T_1$, $T_2$ and $T_3$ are not all equal. Here, $L^{\rm FLCD}(2)$ and $L^{\rm LMYA}(2)$ refer to equations (\ref{eq: L FLCD K3}) and (\ref{eq:fund_limit}), respectively. Note that, when $K=3$ and $r=2$, we obtain that $L^{\rm LMYA}=\frac{1}{2}\left( 1-\frac{2}{3} \right)=\frac{1}{6}$.
 Then, by using the Arithmetic Mean-Geometric Mean (AM-GM) Inequality twice to obtain
 \be
 \label{eq: G 1}
 \frac{T_1 T_2 + T_1 T_3+T_2T_3}{3} \ge \sqrt[3]{(T_1 T_2 T_2)^2},
 \ee
  and
  \be
   \label{eq: G 2}
  \frac{T_1 + T_2 + T_3}{3} \ge \sqrt[3]{T_1 T_2 T_2}.
  \ee
  In both (\ref{eq: G 1}) and (\ref{eq: G 2}), equality holds only when $T_1=T_2=T_3$. By using (\ref{eq: G 1}) and (\ref{eq: G 2}), we can obtain that
 \begin{align}
     & (T_1 T_2 + T_1 T_3+T_2T_3)(T_1 + T_2 + T_3) \notag\\
     & \geq \sqrt[3]{(T_1 T_2 T_2)^2} \cdot \sqrt[3]{T_1 T_2 T_2} 
     =  9T_1T_2T_2.
 \end{align}
Therefore
 \begin{align}
     L^{\rm FLCD}(2) &= \frac{3T_1T_2T_2}{2(T_1 T_2 + T_1 T_3+T_2T_3)(T_1 + T_2 + T_3)} \notag\\
     &\leq \frac{3T_1T_2T_2}{2\cdot 9T_1T_2T_2} 
     = \frac{1}{6},
 \end{align}
 where equality holds only if $T_1=T_2=T_3$. Hence, we complete the proof of Corollary~\ref{corollary: 1}.
 
%

\section{The Proof 
of Theorem 1}
\label{sec: proof thm 1}


Here, we will provide the correctness proof of the general FLCD scheme for $K >3$ and prove the communication-computation tradeoff shown in (\ref{eq: L FLCD}).

In this case, we will first prove (\ref{eq: L FLCD}) in Theorem 1 and then prove the correctness of the FLCD scheme.
To derive the communication load, we will need to count the number of bits transmitted. By the FLCD design, the number of bits in each IV set $\mathcal{V}_n^{k,j}$ is $\lfloor m \rfloor T'_1$ bits. The reason for it is as follows. If $k\in\mathcal{K}_1^i\subseteq\mathcal{K}_1, k \in [K]$, there are  $|\mathcal{K}_1^i|-1=\hat{m}-1=\lfloor m \rfloor$ files that the nodes in $\mathcal{S}_n\setminus k$ have the access to but node $k$ does not. These files are defined by the files mapped to the nodes $\mathcal{S}_n\setminus k$ and a node $k'\in\mathcal{K}_1^i\setminus k$.  Therefore, node $k$ requests $\lfloor m \rfloor$ IVs, each of size $T'_1$ bits, from the nodes of $\mathcal{S}_n\setminus k$. Similarly, if $k\in\mathcal{K}_2^i\subseteq\mathcal{K}_2, k \in [K]$, the number of bits in each IV set $\mathcal{V}_n^{k,j}$ is
$(|\mathcal{K}_2^i|-1)T_2 = (\lfloor m \rfloor - 1)T'_2$ bits 
 that node $k$ requests from the nodes of $\mathcal{S}_n\setminus k$. Since $ \lfloor m \rfloor T'_1= (\lfloor m \rfloor - 1)T'_2$, each IV set $\mathcal{V}_n^{k,j}$ is $\lfloor m \rfloor T'_1$ bits. Consider all shuffle groups $\mathcal{S}_n, n \in [N]$. Each of the $r$ nodes of $\mathcal{S}_n$, 
 sends a message of size $\frac{\lfloor m \rfloor T'_1}{r-1}$ bits. Hence, the communication load is given by
 \if{0}
\begin{align}
    &L^{\rm FLCD} 
    \overset{({\rm a})}{=} \frac{1}{N(K_1T'_1+K_2T'_2)} \cdot N \cdot r \cdot \frac{\lfloor m \rfloor T'_1}{r-1} \notag \\
    & =\frac{1}{r-1} \notag\\
    & \quad \cdot \frac{r\lfloor m \rfloor T'_1}{(\hat{m}K-\lfloor m \rfloor\hat{m}r)T'_1+(\lfloor m \rfloor \hat{m} r -\lfloor m \rfloor K)\cdot\frac{\lfloor m \rfloor}{\lfloor m \rfloor - 1}\cdot T'_1} \notag \\
    &= \frac{1}{r-1} \notag\\
    &\quad \cdot \frac{r\lfloor m \rfloor(\lfloor m \rfloor - 1)}{(\hat{m}K-\lfloor m \rfloor\hat{m}r)(\lfloor m \rfloor - 1)+(\lfloor m \rfloor \hat{m} r -\lfloor m \rfloor K)\lfloor m \rfloor} \notag \\
    &= \frac{1}{r-1} \notag\\
    &\; \cdot \frac{r\lfloor m \rfloor(\lfloor m \rfloor - 1)}{r(\lfloor m \rfloor \hat{m} - \lfloor m \rfloor ^2\hat{m} + \lfloor m \rfloor ^2\hat{m}) + K(\hat{m}(\lfloor m \rfloor -1) - \lfloor m \rfloor^2)} \notag \\
    &= \frac{1}{r-1}\cdot \frac{r\lfloor m \rfloor(\lfloor m \rfloor - 1)}{r\lfloor m \rfloor \hat{m} + K((\lfloor m \rfloor +1)(\lfloor m \rfloor -1) - \lfloor m \rfloor^2)}\notag \\
    &= \frac{1}{r-1}\cdot \frac{r\lfloor m \rfloor(\lfloor m \rfloor - 1)}{r\lfloor m \rfloor \hat{m} - K} \notag \\
    &=\frac{1}{r-1}\left( \frac{\lfloor m \rfloor^2-\lfloor m \rfloor}{\lfloor m \rfloor \hat{m} - m}\right),
\end{align}
\fi
\begin{align}
    &L^{\rm FLCD} 
    \overset{({\rm a})}{=} \frac{1}{N(K_1T'_1+K_2T'_2)} \cdot N \cdot r \cdot \frac{\lfloor m \rfloor T'_1}{r-1} \notag \\
    & 
    =\frac{r\lfloor m \rfloor T'_1/(r-1)}{(\hat{m}K-\lfloor m \rfloor\hat{m}r)T'_1+(\lfloor m \rfloor \hat{m} r -\lfloor m \rfloor K)\cdot\frac{\lfloor m \rfloor}{\lfloor m \rfloor - 1}\cdot T'_1} \notag \\
    &= 
    \frac{r\lfloor m \rfloor(\lfloor m \rfloor - 1)/(r-1)}{(\hat{m}K-\lfloor m \rfloor\hat{m}r)(\lfloor m \rfloor - 1)+(\lfloor m \rfloor \hat{m} r -\lfloor m \rfloor K)\lfloor m \rfloor} \notag \\
    &= 
    \frac{r\lfloor m \rfloor(\lfloor m \rfloor - 1)/(r-1)}{r(\lfloor m \rfloor \hat{m} - \lfloor m \rfloor ^2\hat{m} + \lfloor m \rfloor ^2\hat{m}) + K(\hat{m}(\lfloor m \rfloor -1) - \lfloor m \rfloor^2)} \notag \\
    &= \frac{1}{r-1}\cdot \frac{r\lfloor m \rfloor(\lfloor m \rfloor - 1)}{r\lfloor m \rfloor \hat{m} + K((\lfloor m \rfloor +1)(\lfloor m \rfloor -1) - \lfloor m \rfloor^2)}\notag \\
    &= \frac{1}{r-1}\cdot \frac{r\lfloor m \rfloor(\lfloor m \rfloor - 1)}{r\lfloor m \rfloor \hat{m} - K} \notag \\
    &=\frac{1}{r-1}\left( \frac{\lfloor m \rfloor^2-\lfloor m \rfloor}{\lfloor m \rfloor \hat{m} - m}\right),
\end{align}
where (a) is because $\mathcal{V}_n^{k,j}$ contains $\lfloor m \rfloor T'_1$ bits. Hence, we obtain (\ref{eq: L FLCD}) in Theorem 1.

It remains to prove the correctness of the FLCD scheme when $K>3$. In order to show this, we will need to verify that every node $k$ collects all IVs $V_{k,1},\ldots,v_{k,N}$. This can be seen because node $k$ will receive every IV set $\mathcal{V}_n^k$ for all $n$ such that $k\in\mathcal{S}_n$. Moreover, $\mathcal{V}_n^k$ contains every IV computed by the nodes of $\mathcal{S}_n\setminus k$ but not at node $k$. This includes all IVs from files mapped at nodes $\mathcal{S}_n\setminus k$ 
and at one node from $\mathcal{K}_j^i\setminus k$ where $k\in\mathcal{K}_j^i$. 
By considering all $N$ node groups, this covers all files not available to node $k$. Therefore, node $k$ will receive all requested IVs that are not locally computed.




\section{Proof of Corollary~\ref{corollary: 2}}
\label{sec:proof_cor2}
In this section, we prove the Corollary~\ref{corollary: 2} which states that $L^{\rm FLCD}(r) <\frac{1}{r-1}\left(1-\frac{r}{K}\right)$ when $m$ is not an integer and $K>3$. 
 $L^{\rm FLCD}(r)$ is given in (\ref{eq: L FLCD}).
Assume that $m=\lfloor m \rfloor + a > 1$ and $0<a<1, a \in \mathbb{R}$ such that $m$ is not an integer. 
Then, using the fact that $a-a^2 > 0$, it can be seen that
\begin{align}
    &\lfloor m \rfloor^3 -\lfloor m \rfloor^2 + a\lfloor m \rfloor^2-a\lfloor m \rfloor \notag\\
    &<\lfloor m \rfloor^3 -\lfloor m \rfloor^2 + a\lfloor m \rfloor^2-a\lfloor m \rfloor+a-a^2.
\end{align}
Then, we obtain
\begin{align}
\label{eq: 11}
    (\lfloor m \rfloor+a)(\lfloor m \rfloor^2-\lfloor m \rfloor) &< (\lfloor m \rfloor+a-1)(\lfloor m \rfloor^2-a).
\end{align}
Using the fact that $\lfloor m \rfloor^2\geq1>a$, (\ref{eq: 11}) implies
\begin{align}
    \frac{\lfloor m \rfloor^2-\lfloor m \rfloor}{\lfloor m \rfloor^2-a} &< \frac{\lfloor m \rfloor+a-1}{\lfloor m \rfloor+a},
\end{align}
Since $m=\lfloor m \rfloor + a$, we obtain
\begin{align}
    \frac{\lfloor m \rfloor^2-\lfloor m \rfloor}{\lfloor m \rfloor^2+\lfloor m \rfloor-m} &< \frac{m-1}{m},
\end{align}
which implies
\begin{align}
    \frac{\lfloor m \rfloor^2-\lfloor m \rfloor}{\lfloor m \rfloor(\lfloor m \rfloor+1)-m} &< 1-\frac{1}{m}.
\end{align}
Finally, since $\hat{m}=\lfloor m \rfloor + 1$ and $m=\frac{K}{r}$, we obtain
    \begin{align}
    \frac{\lfloor m \rfloor^2-\lfloor m \rfloor}{\lfloor m \rfloor\hat{m}-m} &< 1-\frac{r}{K}.
    \end{align}
Hence,
    \begin{align}
    L^{\rm FLCD}(r) = \frac{1}{r-1}\left( \frac{\lfloor m \rfloor^2-\lfloor m \rfloor}{\lfloor m \rfloor\hat{m}-m}\right) &< \frac{1}{r-1}\left( 1-\frac{r}{K} \right).
\end{align}
Therefore, we complete the proof of Corollary~\ref{corollary: 2}.

\section*{Acknowledgement} 
We thank Aaron Goh for the help on the Amazon EC2 implementations of the algorithms used in this paper.

\bibliographystyle{IEEEbib}
\bibliography{references_d2d}

\end{document}